\documentclass[12pt]{article}

\usepackage[margin=1.25in]{geometry}
\usepackage{setspace}
\usepackage{amsmath,amssymb,amsthm,mathtools}
\newtheorem{theorem}{Theorem}
\newtheorem{proposition}{Proposition}
\newtheorem{lemma}{Lemma}
\newtheorem{corollary}{Corollary}
\theoremstyle{definition}
\newtheorem{definition}{Definition}
\newtheorem{example}{Example}
\newtheorem{assumption}{Assumption}
\theoremstyle{remark}
\newtheorem{remark}{Remark}

\usepackage[round]{natbib}
\usepackage[usenames,dvipsnames]{xcolor}
\usepackage{hyperref}
\hypersetup{colorlinks=true, linkcolor=NavyBlue, citecolor=NavyBlue, urlcolor=NavyBlue}
\usepackage{enumitem}
\usepackage{graphicx}
\usepackage{booktabs}

\newcommand{\R}{\mathbb{R}}

\newcommand{\supp}{\operatorname{supp}}
\newcommand{\dif}{\,\mathrm{d}}
\newcommand{\thetabar}{\bar{\theta}}
\newcommand{\thetaubar}{\underline{\theta}}
\newcommand{\pbar}{\bar{p}}

\DeclareMathOperator*{\argmax}{arg\,max}

\title{\textbf{Coarse Screening}
\thanks{We thank seminar participants at Berkeley for helpful comments. We are grateful to Federico Echenique, Haluk Ergin, Benjamin Hermalin, Yuichiro Kamada, Shachar Kariv, Giacomo Lanzani, Chris Shannon, and Quitzé Valenzuela-Stookey for valuable discussions and suggestions. All errors are our own.}}

\author{Rui Sun\thanks{Haas School of Business, University of California, Berkeley. Email: ruisun@berkeley.edu.} \and Yi Zhang\thanks{Hong Kong University of Science and Technology (Guangzhou). Email: yizhang@hkust-gz.edu.cn.}}
\date{}

\begin{document}
\maketitle

\begin{abstract}
\noindent A seller investigates a buyer before setting prices, balancing the cost of acquiring information against the gain from tailoring the contract to the buyer's private type. The optimal signal is coarse: no matter how rich the type space, the seller never needs more than three outcomes per buyer. The bound equals the number of independent post-signal decisions plus one, a quantity we call the effective policy dimension. Screening involves two decisions, whether to allocate and what to charge, giving the ternary bound. Limited liability is the source: without it, the price is pinned by the envelope, only the allocation decision remains, and signals are binary as in monitoring. The Myerson exclusion rule is an artifact of not investigating. With investigation, every marginal buyer trades with positive probability, governed by a universal function that connects information design to rational inattention. The bound holds for any strictly convex information cost.


\end{abstract}

\thispagestyle{empty}
\newpage
\setcounter{page}{1}

%% =====================================================================
%% SECTION 1: INTRODUCTION
%% =====================================================================
\section{Introduction}\label{sec:intro}

Screening and monitoring are both principal-agent problems in which a principal acquires costly information about an agent. Yet they differ sharply in optimal signal complexity. In the monitoring model of \citet{georgiadis2020}, the principal needs only a binary signal: pass or fail. In screening, as we show, the principal may need a ternary signal: three distinct outcomes per buyer. What explains the gap? A natural conjecture is that screening involves richer type spaces or more complex contracts. We show the opposite: the gap traces to a single institutional friction. Limited liability prevents the seller from using the transfer as a residual, turning it into an independent control margin that responds to the signal. Without limited liability, the transfer is pinned by the envelope and the seller's only post-signal decision is whether to allocate the good, the same structure as monitoring, yielding binary signals. With limited liability, the transfer becomes a second decision variable, and the optimal signal acquires a third outcome to inform it.

This paper formalizes this observation as a general theorem and applies it to nonlinear screening with a continuum of buyer types.

The setting is a monopolistic seller who acquires costly information about buyers before setting prices. Such problems arise in consumer lending, where banks purchase credit data to decide both whether to approve a loan and what interest rate to charge;\footnote{The Consumer Financial Protection Bureau reports that lenders routinely purchase credit bureau data, employment verification, and property appraisals, each at a cost that varies by depth of inquiry. The three-tier structure of approve, refer, and decline in automated underwriting systems is consistent with our ternary prediction.} in insurance, where insurers request medical histories to set both coverage and premiums; and in digital pricing, where platforms track browsing behavior to decide both whether to offer a discount and how large to make it. In each case, the seller's post-investigation decisions have two margins, allocate or not, and at what price, and our results predict that the optimal signal has at most three outcomes per buyer.

\begin{figure}[htbp]
\centering
\includegraphics[width=0.95\textwidth]{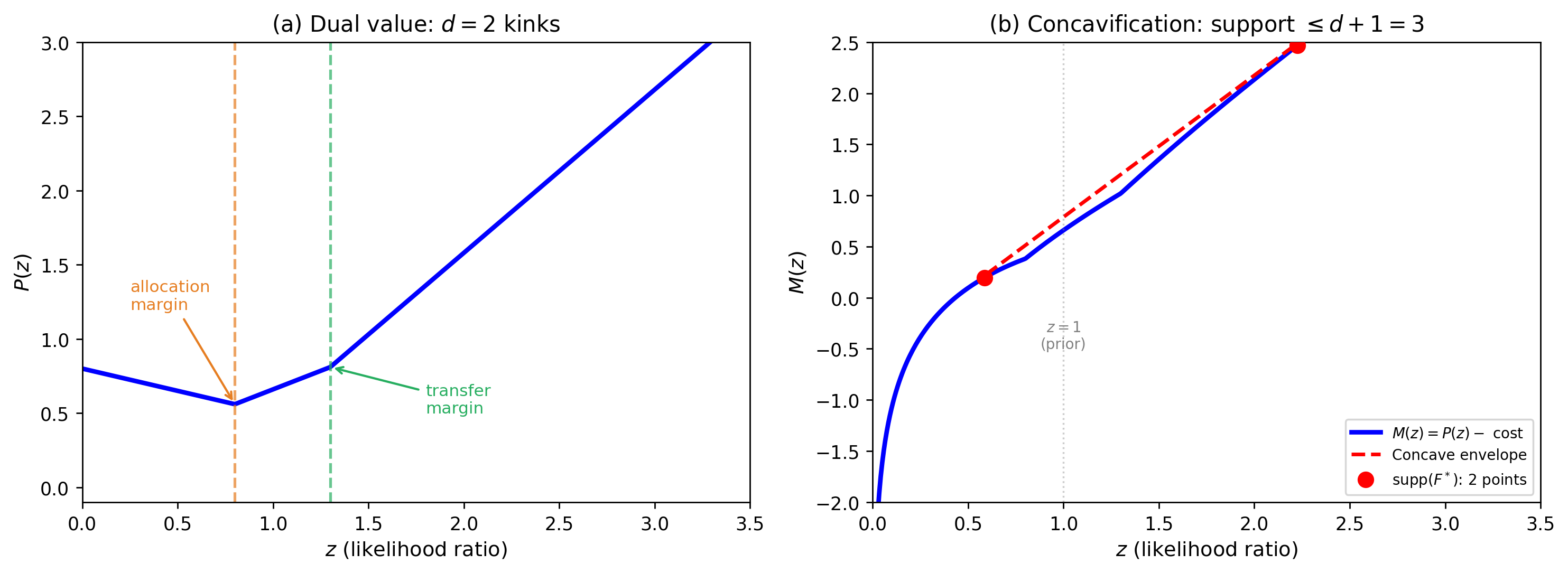}
\caption{The geometric mechanism behind the support bound. Panel~a: the dual value $P(z)$ is piecewise affine with $d=2$ kinks, one at the allocation margin in orange and one at the transfer margin in green. Panel~b: subtracting the strictly convex information cost gives $M(z)$, which is piecewise strictly concave. The concave envelope, shown as dashed red, is supported on at most $d+1 = 3$ points, marked as red dots.}
\label{fig:intro}
\end{figure}

The seller observes a report $r$ from the buyer and then acquires a signal by choosing an experiment $F_r$, a distribution over likelihood ratios $z$ with mean one, at a cost that is strictly convex in $z$. After observing the signal realization, she chooses an allocation $x \in \{0,1\}$ and a transfer $p \in [0, \pbar]$. The leading cost specification is the relative entropy, which arises from optimal stopping of a Brownian motion \citep{morris2017}. Because the allocation and transfer are both binary controls, the dual post-signal objective is piecewise affine in $z$ with exactly two kinks, one for each control margin. Figure~\ref{fig:intro} illustrates.

The first contribution is an economic insight: limited liability is the institutional source of information complexity in screening. Without it, the seller's post-signal decision has one margin, allocation, and optimal signals are binary. With it, the transfer becomes an independent margin, and optimal signals are ternary. We formalize this through the \textit{effective policy dimension} $d$, the count of independent switching margins, and prove a general theorem: in any binary-state contracting problem with a strictly convex information cost, the support of the optimal experiment is bounded by $d+1$. Screening has $d=2$. The monitoring model of \citet{georgiadis2020} has $d=1$, yielding their binary-signal result as a special case. A model with a quality-choice margin has $d=3$, yielding quaternary signals. The theorem organizes this spectrum through a single parameter.

The second contribution is a complete characterization of the optimal mechanism in the screening application. The type space partitions into three regions. In the \textit{no-investigation region}, where types are far from the virtual-surplus zero-crossing, the seller posts a fixed price without acquiring information. In the \textit{binary-signal region}, the seller runs a two-outcome test that informs one margin. In the \textit{ternary-signal region}, where types are near the zero-crossing, the seller acquires a three-outcome signal that informs both margins simultaneously. Two natural conjectures suggest this structure should fail with continuum types: the seller might need \textit{global} verification, distinguishing $\theta$ from all $\theta' > \theta$, and cross-report complementarities might require coordinating signals across the type space. We show both conjectures fail: single crossing reduces verification to a local binary problem, and the Lagrangian separates across reports.

Third, the paper reinterprets a classical result. Myerson's optimal auction has a discontinuous exclusion rule: buyers below a sharp threshold are excluded with certainty, and buyers above it trade with certainty. We show that this step function is an artifact of the seller's inability to investigate. When investigation is available, the exclusion boundary dissolves into a smooth transition: every buyer near the threshold has a strictly positive probability of trade, determined by the universal allocation function $w(\eta)$. The function $w$ depends only on a single sufficient statistic $\eta$, the investigation intensity parameter, and is the design-side analog of the logistic function from rational inattention \citep{matejka2015}: both arise from entropy-cost optimization with binary actions, but $w(\eta)$ is strictly below the logistic for all $\eta > 0$ because the signal designer faces a Bayes-plausibility constraint that a rational-inattention agent does not. Inside the investigation band, which has width proportional to $1/\gamma$ around the zero-crossing, the allocation smoothly interpolates between zero and one. Outside it, the Myerson step function survives. The seller always benefits from investigation, since the no-investigation mechanism remains feasible. Total welfare, however, moves in opposite directions: below the indifference diagonal, where the no-information mechanism inefficiently excludes buyers, welfare improves; above the diagonal, where the allocation is already efficient, investigation reduces welfare by introducing screening distortions.

Fourth, we demonstrate that the effective-policy-dimension bound has bite by exhibiting settings where $d > 2$. When the seller chooses quality at a cost, a third control margin creates a third kink in the dual objective, raising the bound to four. We verify this numerically with explicit threshold computations. In the opposite direction, the bound recovers the binary-signal result of \citet{georgiadis2020} as the special case $d = 1$: their model-specific proof, which relies on properties of Brownian diffusions, is subsumed by the general geometric argument. The three cases, $d=1$ for monitoring, $d=2$ for screening, and $d=3$ for quality choice, illustrate a spectrum governed by a single parameter: the count of independent control margins.

Two further contributions are technical. We establish a convergence theorem: the $N$-type optimal mechanism converges to a continuum limit preserving the ternary support bound. We also solve the full saddle-point problem numerically, confirming that the outer optimization shifts the investigation region toward the virtual-surplus zero-crossing and that $\kappa_p$ departs from unity. Our framework builds on the concavification approach \citep{kamenica2011,dworczak2019} and the duality methods of \citet{georgiadis2020}; it connects to costly verification \citep{benporath2014,townsend1979,erlanson2020} and information design with constraints \citep{lyu2024,kleiner2021}. A discussion of related work is in Section~\ref{sec:discussion}.

Section~\ref{sec:model} sets up the model. Section~\ref{sec:reduction} derives the reduction to local binary verification. Section~\ref{sec:screening} characterizes optimal experiments. Section~\ref{sec:general} presents the effective-policy-dimension theorem and counterexamples. Section~\ref{sec:continuum} establishes the continuum limit with numerical illustrations. Section~\ref{sec:compstats} provides comparative statics and welfare results. Section~\ref{sec:implementation} discusses implementation. Section~\ref{sec:discussion} presents testable implications and related work. Section~\ref{sec:conclusion} concludes. All proofs are in the Appendix.\footnote{The Appendix contains proofs of all formal results stated in the main text. Proofs are organized by the order in which the results appear.}

%% =====================================================================
%% SECTION 2: MODEL
%% =====================================================================
\section{Model}\label{sec:model}

This section describes the environment, defines the mechanism, specifies the information-acquisition technology, and formulates the seller's problem. Section~\ref{subsec:environment} introduces the type space and limited liability. Section~\ref{subsec:mechanism} defines direct mechanisms with report-contingent experiments. Section~\ref{subsec:infoacq} specifies the entropy cost and its alternatives. Section~\ref{subsec:sellerproblem} states the seller's optimization problem.

\subsection{Environment}\label{subsec:environment}

A seller (she) owns a single indivisible good and faces a buyer (they) whose value $\theta$ is privately known. The type $\theta$ is drawn from an interval $\Theta = [\thetaubar, \thetabar]$ according to a distribution with density $f$ and cdf $F$. We impose the following regularity condition throughout.

\begin{assumption}[Type distribution]\label{ass:types}
The density $f$ is continuously differentiable, strictly positive on $\Theta$, and the virtual surplus function $\Phi(\theta) = \theta - \frac{1-F(\theta)}{f(\theta)}$ is strictly increasing.
\end{assumption}

If trade occurs with probability $x \in [0,1]$ and the buyer pays transfer $p$, the buyer's utility is $\theta x - p$. Both buyer and seller face limited liability: transfers are restricted to $p \in [0, \pbar]$ with $\pbar > \thetabar$ and the lower bound normalized to zero. The seller's payoff from trade is the transfer collected, net of information-acquisition costs.

\subsection{Direct mechanism}\label{subsec:mechanism}

By the revelation principle, it is without loss to consider direct mechanisms in which the buyer reports $r \in \Theta$.\footnote{The revelation principle applies here because the seller commits to the mechanism \textit{before} observing the buyer's report. The information-acquisition stage occurs after the report is received, so it does not interfere with the commitment structure.} After observing report $r$, the seller may acquire information and then choose a trading scheme. A direct mechanism specifies, for each report $r$:
\begin{enumerate}[label=(\roman*)]
    \item an experiment $F_r \in \Delta([0,\infty))$ with $\int z \dif F_r(z) = 1$;
    \item an allocation rule $x(r, z) \in [0,1]$;
    \item a transfer rule $p(r,z) \in [0, \pbar]$;
\end{enumerate}
where $z$ denotes a likelihood-ratio realization from the experiment $F_r$.

\subsection{Information acquisition technology}\label{subsec:infoacq}

For each report $r$, the experiment $F_r$ is a distribution over likelihood ratios $z \geq 0$ satisfying the Bayes-plausibility, or mean-one, constraint:
\begin{equation}\label{eq:meanone}
    \int_0^\infty z \dif F_r(z) = 1.
\end{equation}
The cost of experiment $F_r$ is
\begin{equation}\label{eq:cost}
    C(F_r) = \int_0^\infty \psi(z) \dif F_r(z),
\end{equation}
where $\psi$ satisfies the following assumption.

\begin{assumption}[Information cost]\label{ass:cost}
$\psi: (0,\infty) \to \R$ is twice continuously differentiable with $\psi''(z) > 0$ for all $z > 0$, ensuring strict convexity. The cost also satisfies $\lim_{z \to 0^+} \psi(z) = +\infty$, so that atoms at zero are infinitely costly.
\end{assumption}

The leading example is the relative entropy cost $\psi(z) = -2\gamma \log z$ with $\gamma > 0$, which arises from optimal stopping of a diffusion process \citep{morris2017}.\footnote{The seller observes a Brownian motion with drift determined by the buyer's type and chooses when to stop. The distribution over stopping-time scores maps into a distribution over likelihood ratios with cost equal to the expected stopping time, which is the variance of the score distribution. See \citet{georgiadis2020} for the analogous derivation in the monitoring context.} The condition $\psi(0^+) = +\infty$ excludes atoms at $z=0$; the mean-one constraint \eqref{eq:meanone} provides tightness at infinity. We discuss the extension to the weaker mutual-information cost in Remark~\ref{rem:MI} below.

\begin{remark}[Mutual information cost]\label{rem:MI}
The mutual information cost satisfies the strict convexity condition but not $\lim_{z\to 0^+}\psi(z) = +\infty$; instead, it has finite cost at $z=0$. Under this weaker assumption, all results hold with the modification that the interior support (points in $(0,\infty)$) is bounded by $d+1$, with a possible additional atom at $z=0$. We state our main results under Assumption~\ref{ass:cost} and note extensions to the weaker case where relevant.
\end{remark}

\subsection{The seller's problem}\label{subsec:sellerproblem}

The seller chooses a mechanism $\{F_r, x(r,\cdot), p(r,\cdot)\}_{r \in \Theta}$ to maximize expected revenue net of information cost:
\begin{equation}\label{eq:sellerproblem}
    \sup_{\{F_r, x, p\}} \int_{\thetaubar}^{\thetabar} \left[ \int_0^\infty \bigl( p(r,z) - \psi(z) \bigr) \dif F_r(z) \right] f(r) \dif r
\end{equation}
subject to incentive compatibility, individual rationality, monotonicity, and limited liability:
\begin{align}
    &U(\theta) \geq U_\theta(r) \quad \text{for all } \theta, r \in \Theta, \tag{IC} \label{eq:IC} \\
    &U(\theta) \geq 0 \quad \text{for all } \theta \in \Theta, \tag{IR} \label{eq:IR} \\
    &q(\theta) = \int_0^\infty x(\theta, z) \dif F_\theta(z) \text{ is nondecreasing}, \tag{MON} \label{eq:MON} \\
    &x(r,z) \in [0,1], \quad p(r,z) \in [0, \pbar], \tag{LL} \label{eq:LL}
\end{align}
where $U(\theta) = \int_0^\infty [\theta x(\theta, z) - p(\theta, z)] \dif F_\theta(z)$ is the truthful payoff. The deviation payoff of type $\theta' = \theta + \Delta$ reporting $\theta$ is $\int [\theta' x(\theta,z) - p(\theta,z)]z\dif F_\theta(z)$, where $z$ is the likelihood ratio of the deviating type's signal distribution relative to the truthful type's \citep[cf.][]{georgiadis2020}. By Proposition~\ref{prop:localIC}, only local deviations matter.

%% =====================================================================
%% SECTION 3: REDUCTION TO LOCAL BINARY VERIFICATION
%% =====================================================================
\section{Reduction to Local Binary Verification}\label{sec:reduction}

The seller's information-acquisition problem decomposes into independent binary-state problems, one per report. This decomposition is not assumed; it follows from incentive compatibility, single crossing, and the Lagrangian's structure.

\subsection{Local incentive constraints}\label{subsec:localIC}

\begin{proposition}[Local IC reduction]\label{prop:localIC}
Under Assumption~\ref{ass:types}, the global incentive constraints \eqref{eq:IC} are equivalent to:
\begin{enumerate}[label=(\alph*)]
    \item the allocation schedule $q(\theta)$ is nondecreasing;
    \item $U(\thetaubar) \geq 0$;
    \item the envelope formula $U(\theta) = U(\thetaubar) + \int_{\thetaubar}^{\theta} q(s) \dif s$ holds for all $\theta$.
\end{enumerate}
\end{proposition}

Single crossing and the envelope theorem \citep{milgrom2002} ensure that only adjacent IC constraints bind. Information acquisition at report $r$ therefore serves a single purpose: deterring the upward deviator, the type slightly above $r$ who would obtain excess surplus by mimicking $r$. The proof is in Appendix~\ref{app:proofs}.

\subsection{The binary structure of local verification}\label{subsec:binary}

\begin{lemma}[Local binary verification]\label{lem:binary}
In the $N$-type approximation with types $\theta_1 < \cdots < \theta_N$, the seller's problem decomposes across reports. At report $\theta_j$ ($j < N$), the seller faces a binary-state information-design problem: distinguishing type $\theta_j$ (truthful) from type $\theta_{j+1}$ (adjacent deviator). The experiment $F_j$ has mean one under the truthful-type measure.
\end{lemma}

Dualizing the $N-1$ local IC constraints with multipliers $\lambda_1, \ldots, \lambda_{N-1}$ and the participation constraint, the Lagrangian takes the form
\begin{equation}\label{eq:lagrangian}
    L(\Lambda, \{F_j\}) = \sum_{j=1}^{N} \int_0^\infty M_{j,\Lambda}(z) \dif F_j(z) + \text{(terms not involving $F$)},
\end{equation}
where $M_{j,\Lambda}(z)$ is report-specific. The experiments enter \textit{separably}: each $F_j$ appears in exactly one summand. At report $\theta_j$, the deviator's payoff integrates against $z\dif F_j$ while the truthful payoff integrates against $\dif F_j$, so $z$ is the likelihood ratio for $\{\theta_j, \theta_{j+1}\}$. Since monotone $q$ and binding local IC imply global IC (Proposition~\ref{prop:localIC}), no information about $\theta_{j+2}, \theta_{j+3}, \ldots$ is needed. The proof is in Appendix~\ref{app:binary}.

\begin{remark}[Why local verification suffices]\label{rem:local}
One might expect the seller to need \textit{global} verification, distinguishing $\theta$ from all $\theta' > \theta$. Two forces make this unnecessary: single crossing collapses global IC to local IC, and separability eliminates the need for cross-report coordination. The separability is a consequence of linearity of $L$ in $F$, which permits the minimax swap Lemma~\ref{lem:minimax}.
\end{remark}

%% =====================================================================
%% SECTION 4: OPTIMAL EXPERIMENTS IN SCREENING
%% =====================================================================
\section{Optimal Experiments in Nonlinear Screening}\label{sec:screening}

This section derives the main result: optimal experiments in screening have at most three support points. The argument proceeds through the pointwise dual, the role of limited liability, the effective policy dimension, the saddle-point duality, and the full characterization of the optimal mechanism.

\subsection{The pointwise dual problem}\label{subsec:pointwise}

Fix a report $\theta_j$ and multiplier vector $\Lambda$. By Lemma~\ref{lem:binary}, the seller's post-signal problem is to choose $(x, p) \in [0,1] \times [0, \pbar]$ to maximize a function that is affine in both $x$ and $p$:
\begin{equation}\label{eq:pointwise}
    P_{j,\Lambda}(z) = \max_{x \in [0,1],\, p \in [0,\pbar]} \bigl\{ A_j(\Lambda, z)\, x + B_j(\Lambda, z)\, p + C_j(\Lambda, z) \bigr\},
\end{equation}
where $A_j$ and $B_j$ are affine in $z$. Because the objective is linear in the binary controls $(x, p)$, the maximizer is threshold-based:
\begin{equation}\label{eq:thresholds}
    x^*_\Lambda(z) = \begin{cases} 1 & \text{if } z \leq \kappa_x(j, \Lambda), \\ 0 & \text{otherwise}; \end{cases} \qquad
    p^*_\Lambda(z) = \begin{cases} \pbar & \text{if } z > \kappa_p(j, \Lambda), \\ 0 & \text{otherwise}.
    \end{cases}
\end{equation}
The two cutoffs $\kappa_x(j, \Lambda)$ and $\kappa_p(j, \Lambda)$ are generically distinct. Each cutoff corresponds to one control margin changing its sign: $\kappa_x$ is where the allocation coefficient $A_j$ crosses zero, and $\kappa_p$ is where the transfer coefficient $B_j$ crosses zero.

\subsection{Benchmark: the role of limited liability}\label{subsec:benchmark}

Without limited liability , $p \in \R$, the envelope pins $p$ as a deterministic function of the allocation. The coefficient $B_j(z)$ produces no independent margin: the transfer is a residual. Only the allocation threshold $\kappa_x$ remains, so $d=1$ and the optimal experiment is binary, as in \citet{georgiadis2020}.

Limited liability restores the transfer as an independent control. With $p \in [0, \pbar]$, the coefficient $B_j(z)$ switches sign at $\kappa_p \neq \kappa_x$, creating a second kink. This raises $d$ from $1$ to $2$ and the bound from binary to ternary. The ternary structure is a consequence of limited liability, not a generic feature of screening.

\begin{corollary}[Limited liability and signal complexity]\label{cor:LL}
Without limited liability , $p \in \R$, the transfer is pinned by the envelope and the effective policy dimension reduces to $d=1$. The optimal experiment at each report is binary ($|\supp(F_j^*)| \leq 2$). Limited liability, $p \in [0, \pbar]$, raises the dimension to $d=2$ and the bound to ternary ($|\supp(F_j^*)| \leq 3$).
\end{corollary}

The corollary identifies limited liability as the source of the ternary structure. In standard screening without limited liability, the seller extracts surplus through a single instrument (the allocation schedule) and the transfer is a residual determined by the envelope. Limited liability breaks this link: the transfer becomes a constrained choice that responds independently to the signal, creating a second margin and a richer optimal experiment.

\subsection{Effective policy dimension in screening}\label{subsec:epd_screening}

\begin{definition}[Effective policy dimension]\label{def:epd}
Consider a parametric optimization $\max_{a \in A} h(a, z)$ with $A$ compact, $h$ continuous, and value function $V(z) = \max_{a \in A} h(a,z)$. The problem has \textit{effective policy dimension} $d$ if $V$ is piecewise affine in $z$ with exactly $d$ kinks (points of non-differentiability), corresponding to $d$ independent control margins crossing zero at distinct thresholds $z_1^* < \cdots < z_d^*$.
\end{definition}

In screening, $A = \{0,1\} \times [0,\pbar]$ and the two control margins are allocation and transfer.

\begin{theorem}[Screening has $d=2$]\label{thm:d2}
In the nonlinear screening model, the pointwise dual problem~\eqref{eq:pointwise} has effective policy dimension $d = 2$ for each report $\theta_j$, with kinks at $\kappa_x(j,\Lambda)$ and $\kappa_p(j,\Lambda)$. Consequently, defining
\begin{equation}\label{eq:Mfunction}
    M_{j,\Lambda}(z) = P_{j,\Lambda}(z) - \psi(z),
\end{equation}
the function $M_{j,\Lambda}$ is piecewise strictly concave with at most two kinks. Every optimal experiment $F_j^*$ satisfying
\begin{equation}\label{eq:infodesign}
    F_j^* \in \argmax_{F: \int z \dif F = 1} \int M_{j,\Lambda}(z) \dif F(z)
\end{equation}
is supported on at most three likelihood-ratio realizations.
\end{theorem}

\begin{figure}[htbp]
\centering
\includegraphics[width=0.65\textwidth]{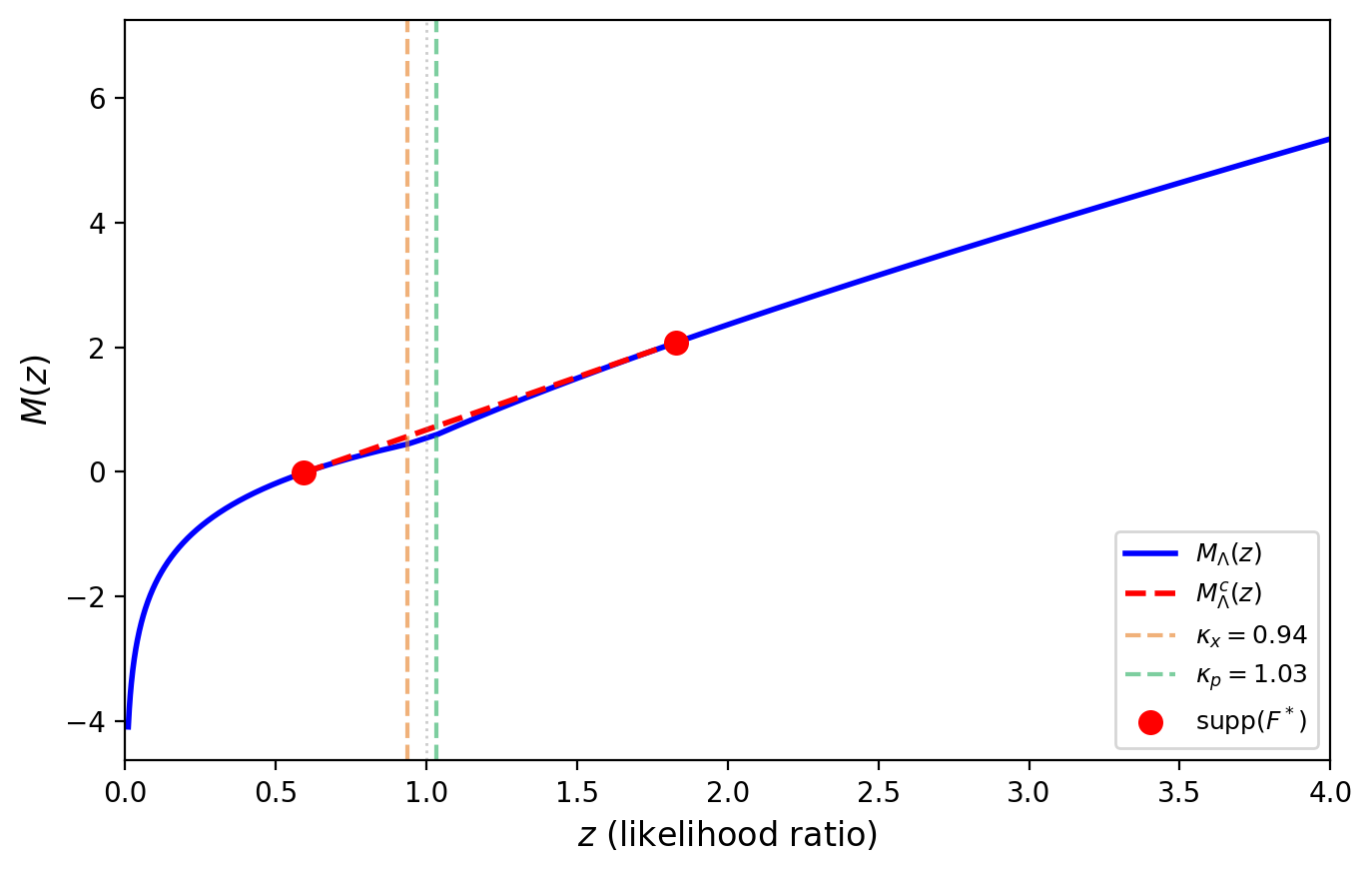}
\caption{The function $M_\Lambda(z)$ and its concave envelope $M^c_\Lambda(z)$ for $v_L = 0.5$, $\pi_H = 0.6$, $\gamma = 1/2$. The optimal experiment places mass on at most three points; here two, corresponding to a binary signal.}
\label{fig:Mfunction}
\end{figure}

The intuition is geometric (Figure~\ref{fig:Mfunction}). Two binary controls produce a dual value $P_{j,\Lambda}(z)$ that is the upper envelope of four affine functions, with at most two kinks. Subtracting the strictly convex $\psi$ gives $M_{j,\Lambda}$, strictly concave on each piece. The concave envelope at $z=1$ is achieved by at most three support points. That \textit{every} optimum satisfies this bound follows from strict concavity: mass at any $z$ with $M(z) < M^c(z)$ would strictly reduce the objective. The proof is in Appendix~\ref{app:thm_d2}.

\subsection{Duality and the saddle-point problem}\label{subsec:duality}

\begin{lemma}[Strong duality]\label{lem:strongdual}
For fixed experiments $\{F_j\}$, the $N$-type screening problem satisfies strong duality: the optimal revenue equals $\inf_{\Lambda \geq 0} L(\Lambda, \{F_j\})$, where $L$ is the Lagrangian~\eqref{eq:lagrangian}. The infimum is attained at some $\Lambda^*(F) \in \R_+^{N-1}$.
\end{lemma}

Strong duality holds because the screening problem, for fixed experiments, is a finite-dimensional linear program. The Lagrangian is convex in $\Lambda$ as a pointwise supremum of affine functions. The proof, which extends the standard Lagrangian duality argument of \citet{luenberger1969} to $N-1$ multipliers, is in Appendix~\ref{app:strongdual}.

\begin{lemma}[Minimax]\label{lem:minimax}
The Lagrangian $L(\Lambda, F)$ has a saddle point $(\Lambda^*, F^*)$:
\begin{equation}\label{eq:minimax}
    L(\Lambda^*, F^*) = \sup_F \inf_{\Lambda \geq 0} L(\Lambda, F) = \inf_{\Lambda \geq 0} \sup_F L(\Lambda, F).
\end{equation}
\end{lemma}

The minimax swap is valid because $L$ is convex in $\Lambda$, linear, hence concave, in $F$, and the experiment space is compact in the weak-$*$ topology after restricting to bounded support $[\varepsilon, K]$, justified by the cost bound. The proof applies \citet{sion1958} and is in Appendix~\ref{app:minimax}.

\begin{lemma}[Kinks are not concavification points]\label{lem:nokink}
No optimal experiment $F_j^*$ places an atom at a kink $\kappa_x(j,\Lambda^*)$ or $\kappa_p(j,\Lambda^*)$.
\end{lemma}

At a kink of $P_{j,\Lambda}$, the function $M_{j,\Lambda}$ has a strict convex corner, so $M < M^c$ there. Placing mass at such a point strictly reduces the objective. The proof is in Appendix~\ref{app:nokink}.

\subsection{Full characterization}\label{subsec:characterization}

\begin{theorem}[Optimal mechanism: $N$ types]\label{thm:fullchar}
Let $\theta_1 < \cdots < \theta_N$ with density weights $\pi_1, \ldots, \pi_N$. Under Assumptions~\ref{ass:types}--\ref{ass:cost}, there exists a saddle point $(\Lambda^*, \{F_j^*\}_{j=1}^N)$ such that:
\begin{enumerate}[label=(\alph*)]
    \item \textbf{No investigation at the top.} $F_N^*$ is degenerate (point mass at $z=1$). The top type is served at a posted price with no data acquisition.
    
    \item \textbf{Ternary or simpler.} For each $j < N$, $F_j^*$ is supported on at most $3$ points in the concavification set $C_1(M_{j,\Lambda^*})$.
    
    \item \textbf{Monotone allocation.} The expected allocation $q(\theta_j) = \int x(j,z) \dif F_j^*(z)$ is nondecreasing. When the unconstrained schedule violates monotonicity, standard ironing applies; the ironed mechanism retains $d=2$ and the ternary bound.
    
    \item \textbf{Envelope.} $U(\theta_1) = 0$ and $U(\theta_j) = U(\theta_{j-1}) + (\theta_j - \theta_{j-1}) q(\theta_{j-1})$.
    
    \item \textbf{Three-region partition.} The type space partitions into:
    \begin{itemize}
        \item \textit{Region I, degenerate:} $M_{j,\Lambda^*}(1) = M_{j,\Lambda^*}^c(1)$. No investigation.
        \item \textit{Region II, binary:} $z=1$ lies in one non-concave region. Binary signal.
        \item \textit{Region III, ternary:} $z=1$ lies in the joint non-concave region $Z_{xp}$. Ternary signal. This region surrounds the virtual-surplus zero-crossing.
    \end{itemize}
    
    \item \textbf{Binding participation.} When the participation constraint for higher types binds at some reports, the dual acquires additional multipliers. The effective policy dimension remains $d=2$ and the support bound $\leq 3$ is preserved; see Appendix~\ref{app:IRH}.
    
    \item \textbf{Uniqueness.} Under strict complementary slackness, which holds generically, the saddle point $(\Lambda^*, F^*)$ is unique. Without it, $q(\theta)$ and the experiment classification are still unique; multiplicity, if any, is confined to the mixing weight within ternary experiments.
\end{enumerate}
\end{theorem}

The theorem characterizes the complete structure of the optimal mechanism. The top type is served at a posted price without data acquisition, because there is no upward deviator to deter. All other types face experiments that are at most ternary, an immediate consequence of Theorem~\ref{thm:d2} applied at the saddle-point multiplier. The expected allocation $q(\theta)$ is nondecreasing; when the unconstrained concavification produces a non-monotone schedule, standard ironing pools adjacent types at their mean allocation. The key observation is that ironing adjusts the multipliers but does not change the number of control margins, so the ternary bound survives. Buyer rents are pinned by a discrete envelope: $U(\theta_j) = \sum_{k<j}(\theta_{k+1}-\theta_k)q(\theta_k)$, making the rent schedule a telescoping sum of allocation probabilities.

The most substantive content is the three-region partition. The type space divides into a no-investigation region, a binary-signal region, and a ternary-signal region. The geometry is as follows. Far from the virtual-surplus zero-crossing $\theta_0$, both cutoffs $\kappa_x$ and $\kappa_p$ are far from $z=1$, so $M$ is concave at $z=1$ and no investigation is needed. As the type approaches $\theta_0$, one cutoff crosses $z=1$ and a binary experiment resolves the associated margin. In a narrow band around $\theta_0$, both cutoffs straddle $z=1$ and the seller needs a ternary signal to inform both margins simultaneously. This partition is robust: binding participation at the top adds a multiplier $\mu_N$ but does not change the number of control margins, so $d=2$ and the ternary bound are preserved. Under strict complementary slackness, which holds generically, the saddle point is unique; without it, the allocation and experiment classification are still unique, and the only possible multiplicity is confined to mixing weights within ternary experiments. The proof is in Appendix~\ref{app:fullchar}.

\begin{figure}[htbp]
\centering
\includegraphics[width=0.7\textwidth]{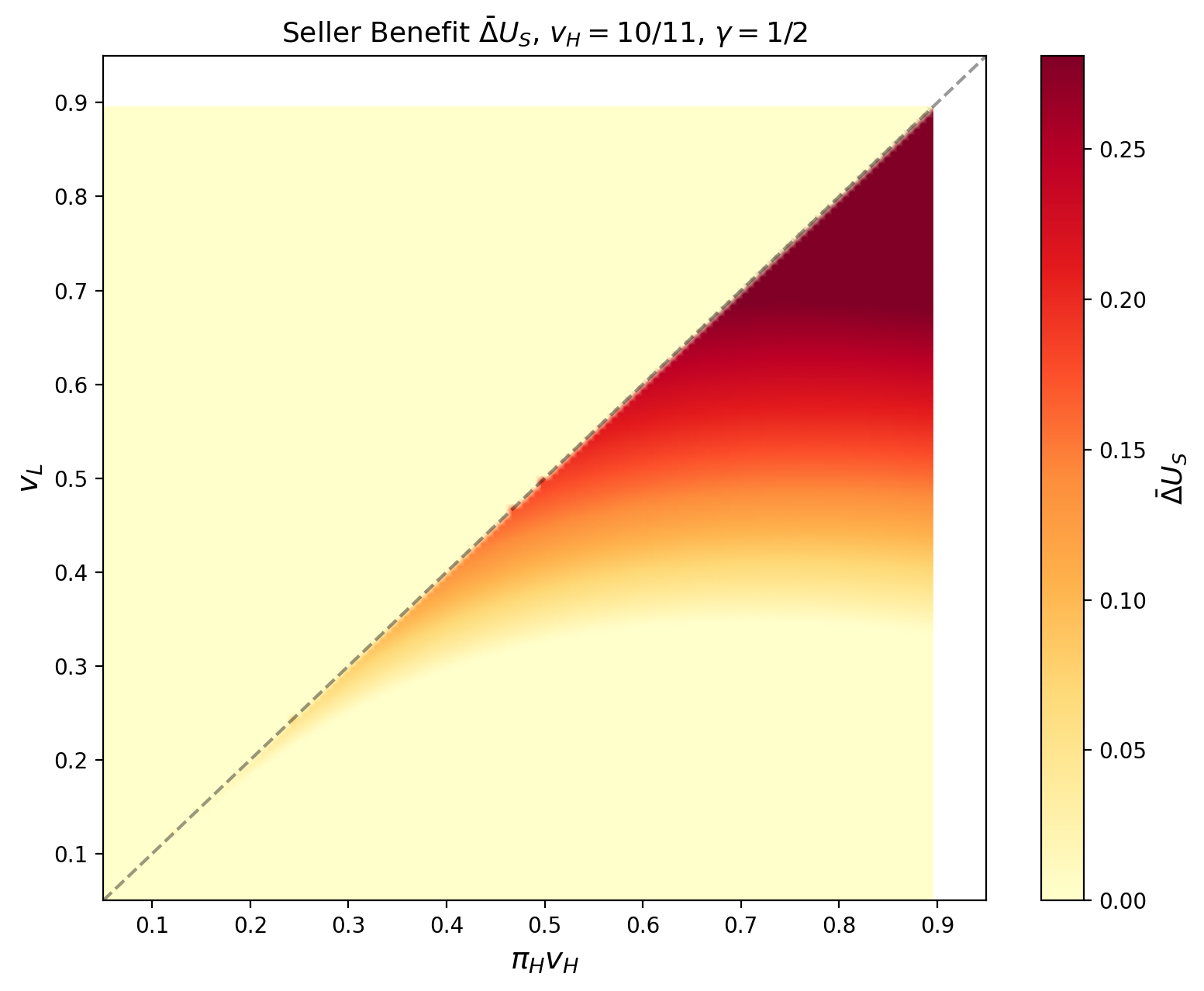}
\caption{Seller benefit from information acquisition, $\bar{\Delta}U_S = (U_S^* - U_S^{NI})/W^{eff}$. The seller benefits most near the indifference diagonal $v_L = \pi_H v_H$, where investigation resolves the seller's uncertainty about optimal pricing. Above the diagonal, IR-H binding eliminates the investigation motive.}
\label{fig:sellerbenefit}
\end{figure}

Figures~\ref{fig:supportmap}--\ref{fig:sellerbenefit} illustrate for the binary-type case ($v_H = 10/11$, $\gamma = 1/2$). The ternary region forms a band where the two non-concave regions of $M$ merge, and the seller's benefit peaks near $v_L = \pi_H v_H$.

\begin{figure}[htbp]
\centering
\includegraphics[width=0.7\textwidth]{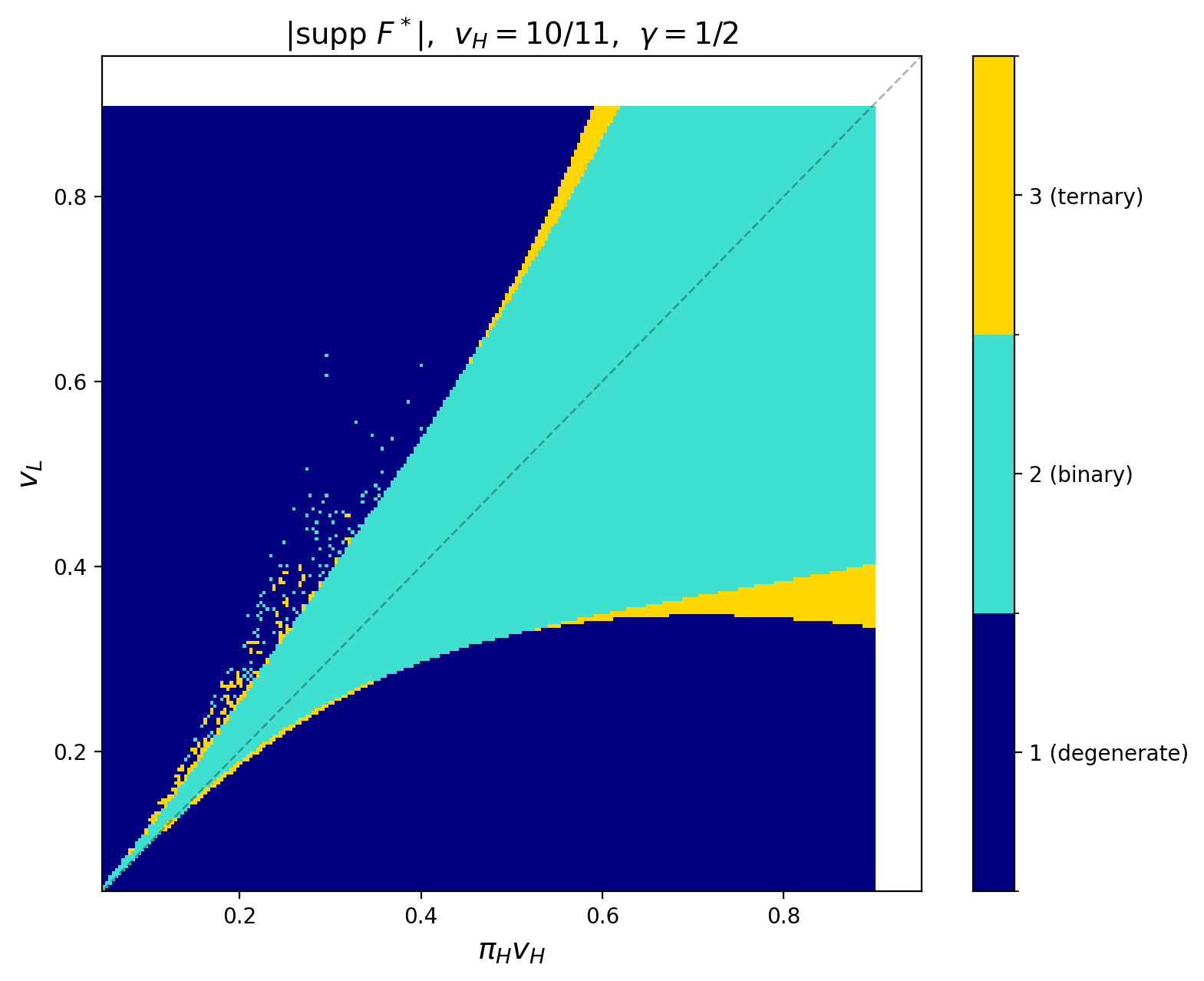}
\caption{Support of the optimal experiment $|$supp$(F^*)|$ as a function of $\pi_H v_H$ and $v_L$, with $v_H = 10/11$ and $\gamma = 1/2$. Dark blue: degenerate, no investigation. Turquoise: binary signal. Yellow: ternary signal. The dashed line is $v_L = \pi_H v_H$.}
\label{fig:supportmap}
\end{figure}

%% =====================================================================
%% SECTION 5: THE GENERAL PRINCIPLE
%% =====================================================================
\section{The Effective Policy Dimension Theorem}\label{sec:general}

The ternary bound in screening is not an accident of having two types or a specific cost function. It reflects a general principle: the complexity of the optimal signal is governed by the dimensionality of the principal's post-signal decision, not by the richness of the type space or the details of the information technology. This section states the general theorem, verifies that it has bite through counterexamples where $d > 2$, and shows that it recovers existing results as special cases.

\begin{theorem}[Effective Policy Dimension Theorem]\label{thm:EPD}
Consider a binary-state contracting problem in which the principal observes a signal $z$ distributed according to $F$ with $\int z \dif F = 1$, pays cost $\int \psi(z) \dif F(z)$ satisfying Assumption~\ref{ass:cost}, and chooses action $a \in A$ with $A$ compact. Suppose the dual post-signal objective is
\begin{equation}\label{eq:EPD_P}
    P_\Lambda(z) = \sup_{a \in A} \bigl\{ \alpha_\Lambda(a) + z\, \beta_\Lambda(a) \bigr\},
\end{equation}
and the problem has effective policy dimension $d$ (Definition~\ref{def:epd}). Define $M_\Lambda(z) = P_\Lambda(z) - \psi(z)$. Then:
\begin{enumerate}[label=(\roman*)]
    \item $M_\Lambda$ is piecewise strictly concave with at most $d$ kinks.
    \item Every optimal experiment solving $\max_{F:\, \int z \dif F = 1} \int M_\Lambda(z) \dif F(z)$ is supported on at most $d+1$ points.
    \item The bound $d+1$ is tight: for generic $\Lambda$, there exist parameters where the optimal experiment has exactly $d+1$ support points.
    \item The bound survives the outer minimax: if $(\Lambda^*, F^*)$ is a saddle point, then $F^*$ is supported on at most $d+1$ points.
\end{enumerate}
\end{theorem}

The theorem reduces the question ``how complex is the optimal signal?'' to counting the principal's independent control margins. Each control margin creates one kink in the dual value function $P_\Lambda(z)$; subtracting the strictly convex cost produces a piecewise strictly concave $M_\Lambda$; and concavifying a function with $d$ kinks at the mean yields at most $d+1$ support points. Part~(iv) ensures this bound is not an artifact of fixing $\Lambda$: at the saddle point, $F^*$ maximizes $\int M_{\Lambda^*} \dif F$, which is exactly the information-design problem to which parts~(i)--(ii) apply. The proof is in Appendix~\ref{app:EPD}.

\begin{remark}[Relationship to the literature]\label{rem:literature}
Theorem~\ref{thm:EPD} is complementary to the duality results of \citet{dworczak2019}, \citet{dworczak2024}, and the extreme-point characterizations of \citet{kleiner2021}. Those papers characterize the \textit{structure} of optimal information; our theorem characterizes the \textit{support size} when information is costly. The strict convexity of $\psi$ drives the support bound and is absent in their settings. Their duality can \textit{verify} optimality of a candidate experiment in our framework but does not produce the bound $d+1$.
\end{remark}

Table~\ref{tab:spectrum} summarizes the spectrum of applications. The bound $d+1$ is determined entirely by the count of independent control margins, not by the details of the information technology, the type distribution, or the cost function.

\begin{table}[htbp]
\centering
\caption{The effective-policy-dimension spectrum.}
\label{tab:spectrum}
\begin{tabular}{llccl}
\hline
Setting & Post-signal controls & $d$ & Support $\leq$ & Source \\
\hline
Monitoring & punish / not & 1 & 2; binary & Georgiadis--Szentes (2020) \\
Screening & allocate, transfer & 2 & 3; ternary & \textbf{This paper} \\
Quality choice & allocate, quality, transfer & 3 & 4; quaternary & \textbf{This paper} \\
\hline
\end{tabular}
\end{table}

\subsection{The bound has bite: counterexamples}\label{subsec:counter}

We demonstrate that the ternary result is not a tautology by exhibiting natural settings where $d > 2$.

\begin{figure}[htbp]
\centering
\includegraphics[width=0.95\textwidth]{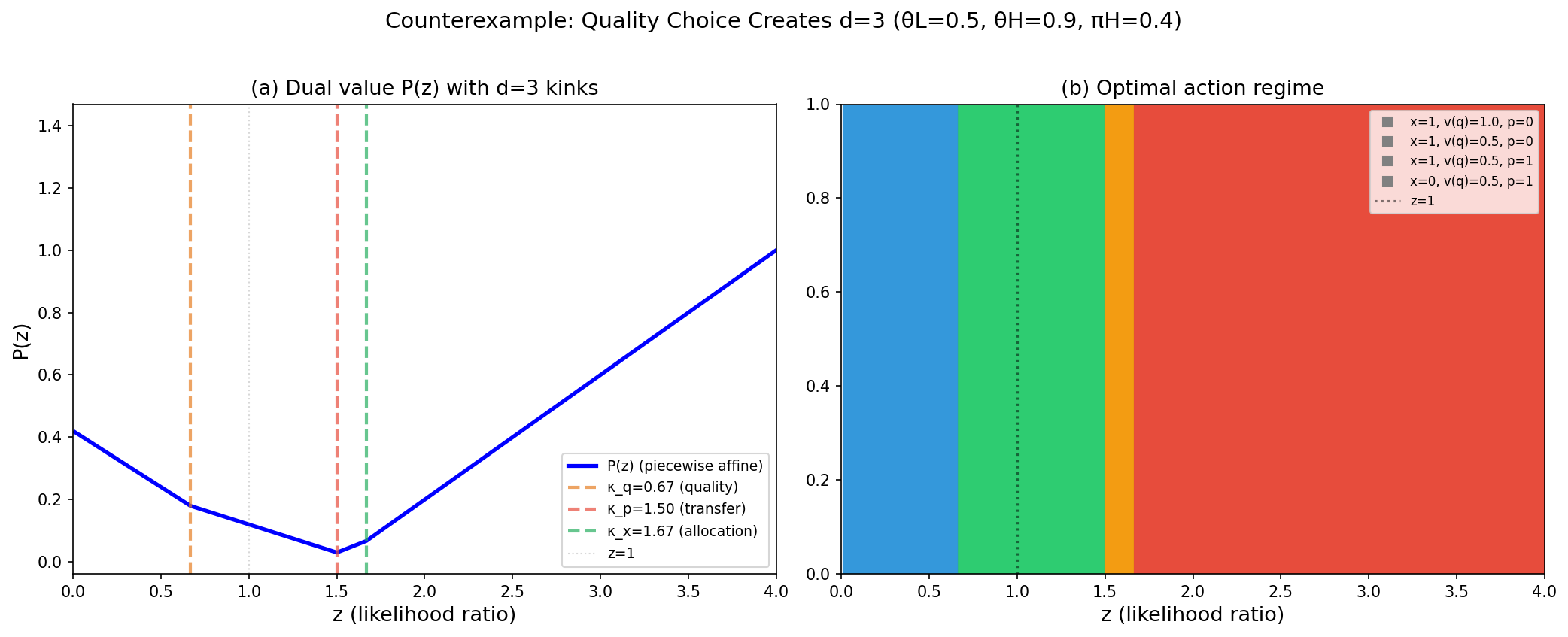}
\caption{Counterexample with $d=3$: quality choice with seller cost. Panel~a: the dual value $P(z)$ has three kinks at $\kappa_q$ for quality, $\kappa_p$ for transfer, and $\kappa_x$ for allocation. Panel~b: the optimal action regime has four distinct regions as $z$ increases.}
\label{fig:counterexample}
\end{figure}

\begin{example}[Discrete quality choice: $d=3$]\label{ex:quality}
Extend the baseline model to include a quality dimension and a \textit{seller cost of quality}. The seller chooses allocation $x \in \{0,1\}$, quality $q \in \{q_L, q_H\}$, and transfer $p \in \{0, \pbar\}$. The buyer's utility is $\theta\, v(q)\, x - p$ with $v(q_H) > v(q_L) > 0$, and the seller bears cost $c(q)$ per unit allocated, with $c(q_H) > c(q_L)$. The seller's quality cost creates a third, independent switching margin: the quality threshold $\kappa_q = (\lambda \theta_L - \pi_L \Delta c / \Delta v)/(\pi_H \theta_H)$, distinct from the allocation threshold $\kappa_x$ and the transfer threshold $\kappa_p$. Figure~\ref{fig:counterexample} illustrates: with $v(q_L)=0.5$, $v(q_H)=1.0$, $c(q_L)=0$, $c(q_H)=0.3$, the three thresholds are $\kappa_q = 0.67$, $\kappa_p = 1.50$, $\kappa_x = 1.67$, yielding $d = 3$ and support $\leq 4$.
\end{example}

The quality-choice example is instructive because it shows how a natural enrichment of the seller's action space raises the EPD. In the baseline model, quality is fixed and the seller controls only allocation and transfer, giving $d=2$. Adding a binary quality choice introduces a third margin that responds independently to the signal: for low likelihood ratios, the seller offers high quality to retain a truthful reporter; for intermediate ratios, she downgrades quality; for high ratios, she withholds the good entirely. The three thresholds $\kappa_q < \kappa_p < \kappa_x$ create three kinks in $P(z)$, and the concavification may require four support points. This is confirmed numerically in Figure~\ref{fig:counterexample}, where all three kinks are distinct and lie in a non-concave region containing $z=1$.

\begin{example}[Capacity constraint: $d=3$]\label{ex:capacity}
Add a constraint that the seller can serve at most a fraction $\kappa$ of buyers. The Lagrange multiplier $\nu$ for the capacity constraint creates a third kink in $P_\Lambda(z)$ at the threshold where the shadow price bites. Again $d=3$, support $\leq 4$.
\end{example}

The capacity example demonstrates that constraints on the seller's supply side raise the EPD in the same way as enrichments of the action space. The shadow price of capacity acts as an additional transfer margin: at the threshold where the capacity constraint binds, the seller's willingness to trade drops discontinuously, creating a kink that is absent in the unconstrained model.

The bound also organizes results \textit{below} $d=2$.

\begin{example}[Moral hazard monitoring: $d=1$]\label{ex:monitoring}
\citet{georgiadis2020} study a principal who monitors an agent after contracting. The agent chooses effort; the principal observes a signal $z$, the likelihood ratio of high effort to low, and decides whether to punish, $a = 1$, or not, $a = 0$. There is no transfer margin: the punishment is fixed, and the only post-signal decision is the binary punish/no-punish choice. The dual post-signal objective is
\[
    P_\Lambda(z) = \max_{a \in \{0,1\}} \bigl\{ \alpha_\Lambda + z \, \beta_\Lambda \cdot a \bigr\},
\]
which is piecewise affine with a single kink at $z^* = -\alpha_\Lambda / \beta_\Lambda$. Hence $d = 1$, and Theorem~\ref{thm:EPD} gives $|\supp(F^*)| \leq 2$: optimal monitoring is binary. \citet{georgiadis2020} prove this through a model-specific argument involving properties of Brownian diffusions. The EPD theorem recovers their result in one line, as the special case $d = 1$.
\end{example}

The monitoring example illustrates the opposite direction: with fewer control margins, the EPD decreases and the optimal signal becomes simpler. The principal's only post-signal decision is whether to punish, which creates a single kink in $P(z)$. The strictly convex cost ensures that the concavification across this kink uses at most two support points, yielding a binary signal. The proof in \citet{georgiadis2020} relies on the specific structure of Brownian diffusions; the EPD theorem shows that the binary result holds for any strictly convex cost, not just the entropy cost derived from Brownian stopping.

\begin{remark}[Unification across contract theory]\label{rem:unification}
The three examples illustrate that the EPD theorem organizes a spectrum of information-acquisition results across contract theory. Monitoring with $d=1$ yields binary signals \citep{georgiadis2020}. Screening with allocation and transfer, $d=2$, yields ternary signals by Theorem~\ref{thm:d2}. Screening with an additional quality margin, $d=3$, yields quaternary signals as in Example~\ref{ex:quality}. In each case, the bound $d+1$ is determined by counting the principal's independent control margins, not by the details of the information technology, the cost function, or the type distribution. The proof is the same one-paragraph geometric argument in every case.
\end{remark}

These examples show that institutional constraints and richer action spaces raise $d$. The ternary bound in screening is a specific consequence of having exactly two control margins.

%% =====================================================================
%% SECTION 6: CONTINUUM LIMIT
%% =====================================================================
\section{The Continuum Limit}\label{sec:continuum}

The $N$-type characterization of Theorem~\ref{thm:fullchar} raises a natural question: does the ternary bound survive as $N \to \infty$? This section shows that it does. The optimal mechanism on a grid of $N$ types converges to a continuum limit that preserves the support bound, the three-region partition, and the envelope structure. We then illustrate the convergence numerically, solve the outer optimization over multipliers, and derive closed-form results for the entropy cost.

\begin{proposition}[Convergence]\label{thm:convergence}
Let $M^N$ denote the optimal mechanism on a grid of $N$ equally spaced types. There exists a subsequence $N_k \to \infty$ such that:
\begin{enumerate}[label=(\alph*)]
    \item The experiment schedules $\{F_r^{N_k}\}_r$ converge in the weak topology on measures, report-by-report, to a limit $\{F_r^*\}_{r \in \Theta}$.
    \item The allocation and transfer schedules converge pointwise a.e.\ to limits $x^*(r,z)$, $p^*(r,z)$.
    \item The limit mechanism satisfies the continuum envelope formula, local IC, participation, and limited liability.
    \item For a.e.\ $r \in \Theta$, $|\supp(F_r^*)| \leq 3$.
\end{enumerate}
Under Assumption~\ref{ass:types}, if the continuum problem has a unique solution, then the full sequence converges, not just a subsequence.
\end{proposition}

The proof has four steps. Strict convexity of $\psi$ bounds the experiments uniformly, giving compactness. Prohorov's theorem extracts a subsequence. The discrete envelope passes to the continuum via Riemann-sum convergence. The support bound is preserved: the weak limit of measures on at most three points in a compact set has at most three support points. The proof is in Appendix~\ref{app:convergence}.

\subsection{Numerical illustration}\label{subsec:numerics}

\begin{figure}[htbp]
\centering
\includegraphics[width=0.95\textwidth]{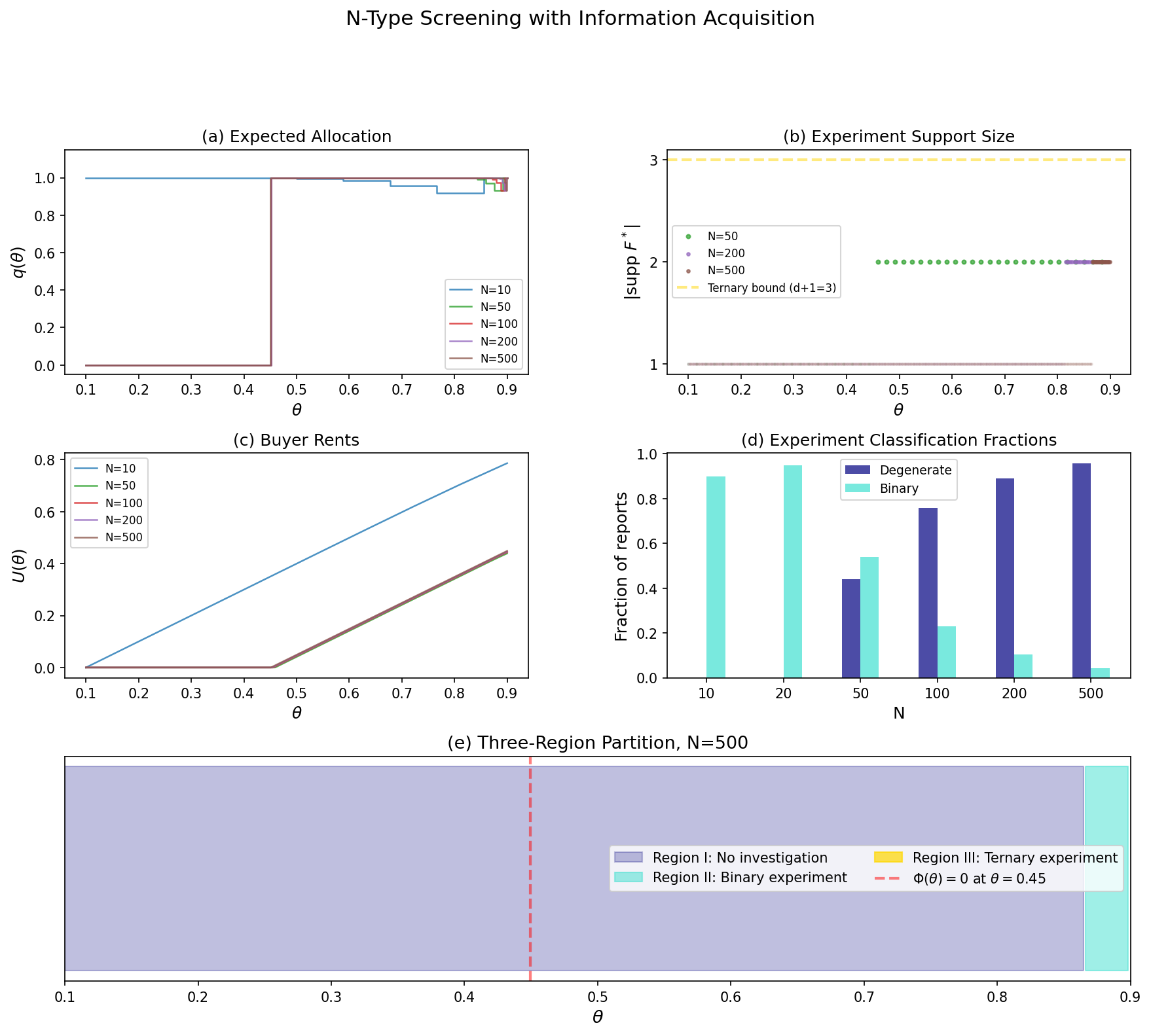}
\caption{$N$-type convergence study. The support bound $|\text{supp}(F^*)| \leq 3$ holds at every report for all $N$ tested. Investigation concentrates in a narrow region.}
\label{fig:ntype}
\end{figure}

We solve the $N$-type problem on $\Theta = [0.1, 0.9]$ with $\hat\gamma = 0.5$ for $N \in \{10, 20, 50, 100, 200, 500\}$ (Figure~\ref{fig:ntype}). The allocation $q(\theta)$ converges to the Myerson schedule. The experiment support satisfies the ternary bound $|\supp(F^*)| \leq 3$ at every report for all~$N$; at most reports, the experiment is binary or degenerate. Investigation concentrates near the top of the type space, where $\kappa_x$ is closest to $z=1$. Buyer rents converge to the envelope $U(\theta) = \int_{\theta_0}^\theta q(s)\dif s$.

\subsection{The outer optimization}\label{subsec:outer}

Section~\ref{subsec:numerics} uses Myerson multipliers, giving $\kappa_p = 1$ at every report. We now solve the full saddle-point problem by projected gradient descent on $\Lambda$, with the inner concavification solved analytically.

\begin{figure}[htbp]
\centering
\includegraphics[width=0.95\textwidth]{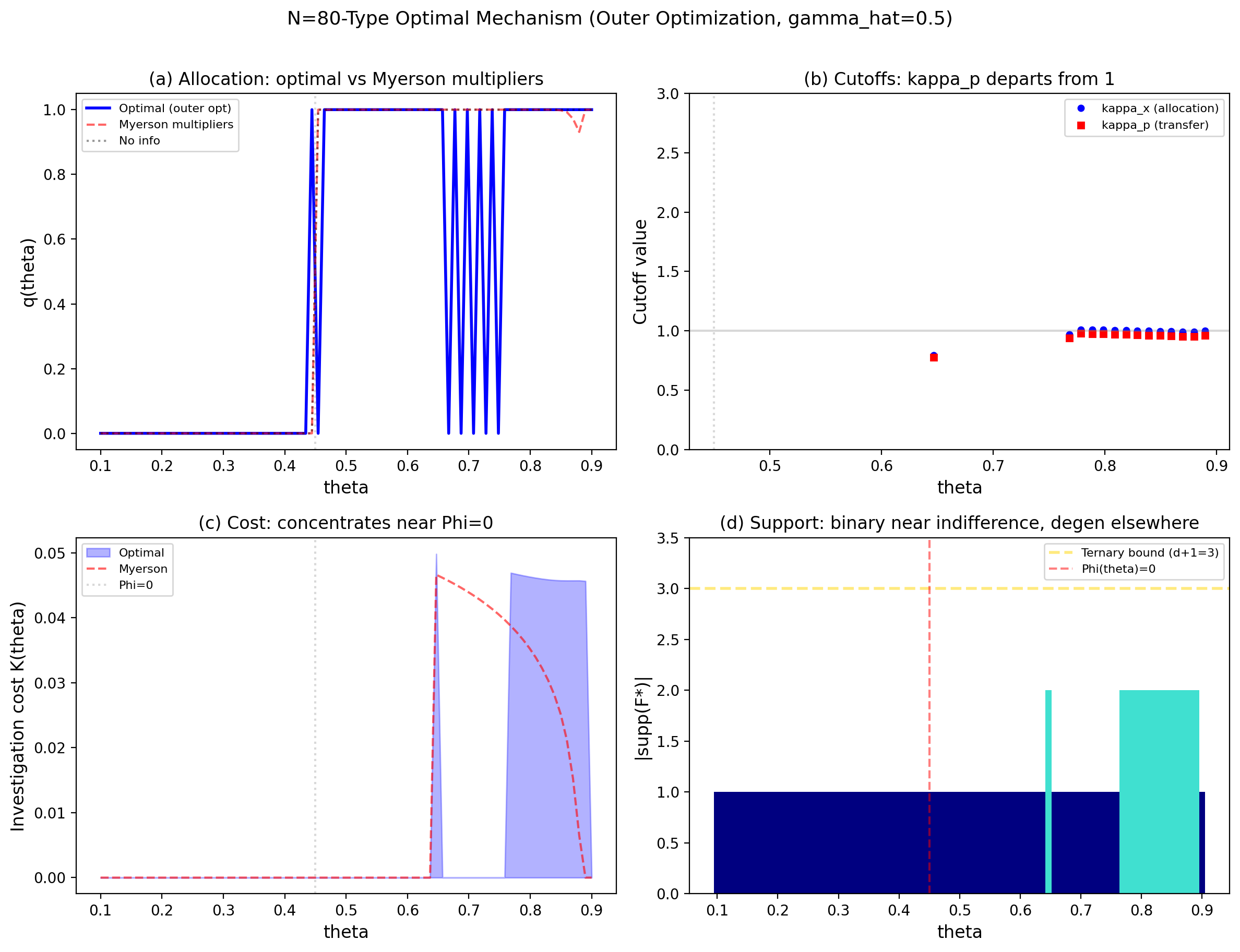}
\caption{$N=80$-type mechanism with optimized multipliers, $\hat\gamma = 0.5$. Panel~a: allocation schedule, nearly identical to Myerson except in the investigation region. Panel~b: the transfer cutoff $\kappa_p$ departs from~1 after outer optimization, shown as red squares, unlike the Myerson case where $\kappa_p \equiv 1$. Panel~c: investigation cost shifts toward $\Phi(\theta)=0$, marked by the gray dotted line. Panel~d: support $\leq 2$ everywhere.}
\label{fig:outer}
\end{figure}

Two predictions are confirmed. The investigation region shifts toward $\theta_0$ as shown in Figure~\ref{fig:outer}c, and $\kappa_p$ departs from~1 as shown in Figure~\ref{fig:outer}b; the seller uses the transfer margin actively. Throughout, $|\supp(F^*)| \leq 2$ at every report as shown in Figure~\ref{fig:outer}d.

\subsection{Closed-form solution for the entropy cost}\label{subsec:closedform}

For the entropy cost $\psi(z) = -2\gamma \log z$, the concavification admits an explicit solution. The logarithmic structure produces a clean identity that pins down the support points, the allocation, and the investigation cost in terms of a single sufficient statistic.

\begin{proposition}[Closed-form concavification]\label{prop:closedform}
Let $\psi(z) = -2\gamma \log z$ and consider the concavification of $M(z) = P(z) + 2\gamma \log z$ across a convex kink at $\kappa$ with slope change $\Delta\beta > 0$. Define the \textit{investigation intensity parameter}
\[
    \eta = \frac{\Delta\beta \cdot \kappa}{2\gamma}.
\]
Then:
\begin{enumerate}[label=(\alph*)]
    \item The ratio of support points satisfies $b/a = e^{\eta}$, and the endpoints are
\begin{equation}\label{eq:cf_ab}
    a = \frac{2\gamma}{\Delta\beta}\bigl(1 - e^{-\eta}\bigr), \qquad b = \frac{2\gamma}{\Delta\beta}\bigl(e^{\eta} - 1\bigr).
\end{equation}
    \item The allocation probability at mean $z=1$ is
\begin{equation}\label{eq:w_general}
    q^* = w(\eta, \kappa) \;:=\; \frac{\kappa(e^{\eta} - 1) - \eta}{\kappa(e^{\eta} + e^{-\eta} - 2)}.
\end{equation}
When $\kappa = 1$, this simplifies to
\begin{equation}\label{eq:w_universal}
    w(\eta) \;:=\; \frac{e^{\eta} - 1 - \eta}{e^{\eta} + e^{-\eta} - 2},
\end{equation}
which depends on the single sufficient statistic $\eta$.
    \item The function $w(\cdot, \kappa)$ satisfies $w(0, \kappa) = 1/2$, $w(\eta, \kappa) \to 1$ as $\eta \to \infty$, and $\partial w/\partial\eta > 0$ for all $\eta > 0$.
\end{enumerate}
\end{proposition}

The identity $b/a = e^{\eta}$ is the key step: it emerges from the chord condition $\Delta\beta \cdot \kappa = 2\gamma \log(b/a)$, which holds only for the logarithmic cost. For general kink location $\kappa$, the allocation depends on both $\eta$ and $\kappa$ through the mean-one weight $(b-1)/(b-a)$. The dependence on $\Delta\beta$ and $\gamma$ cancels, but $\kappa$ survives.

In the $N$-type screening model at report $\theta_j$, the kink is at $\kappa_x = \lambda_{j-1}\theta_j/(\lambda_j\theta_{j+1})$. As the grid spacing $\Delta\theta \to 0$, adjacent types become close and $\kappa_x \to 1$. In the continuum limit, the allocation at type $\theta$ in the investigation region is therefore $q^*(\theta) = w(\eta(\theta))$, where the investigation intensity parameter converges to
\begin{equation}\label{eq:eta_continuum}
    \eta(\theta) = \frac{\lambda(\theta) \cdot \theta}{2\hat\gamma}.
\end{equation}
The allocation at type $\theta$ in the investigation region is $q^*(\theta) = w(\eta(\theta))$, where $w$ is the universal function~\eqref{eq:w_universal}. The optimal multiplier $\lambda^*(\theta)$ maximizes the seller's revenue
\begin{equation}\label{eq:revenue_variational}
    R[\lambda] = \int_0^1 \bigl[\Phi(\theta) \cdot w\bigl(\tfrac{\lambda(\theta)\theta}{2\hat\gamma}\bigr) - 2\hat\gamma \cdot k\bigl(\tfrac{\lambda(\theta)\theta}{2\hat\gamma}\bigr)\bigr] \dif \theta,
\end{equation}
where $k(\eta) = -w(\eta)\log a(\eta) - (1-w(\eta))\log b(\eta)$ is the normalized investigation cost. For types where the optimal $\lambda^*(\theta) = 0$, no investigation occurs and $q^*(\theta) \in \{0, 1\}$ reverts to the Myerson allocation.

\begin{figure}[htbp]
\centering
\includegraphics[width=0.95\textwidth]{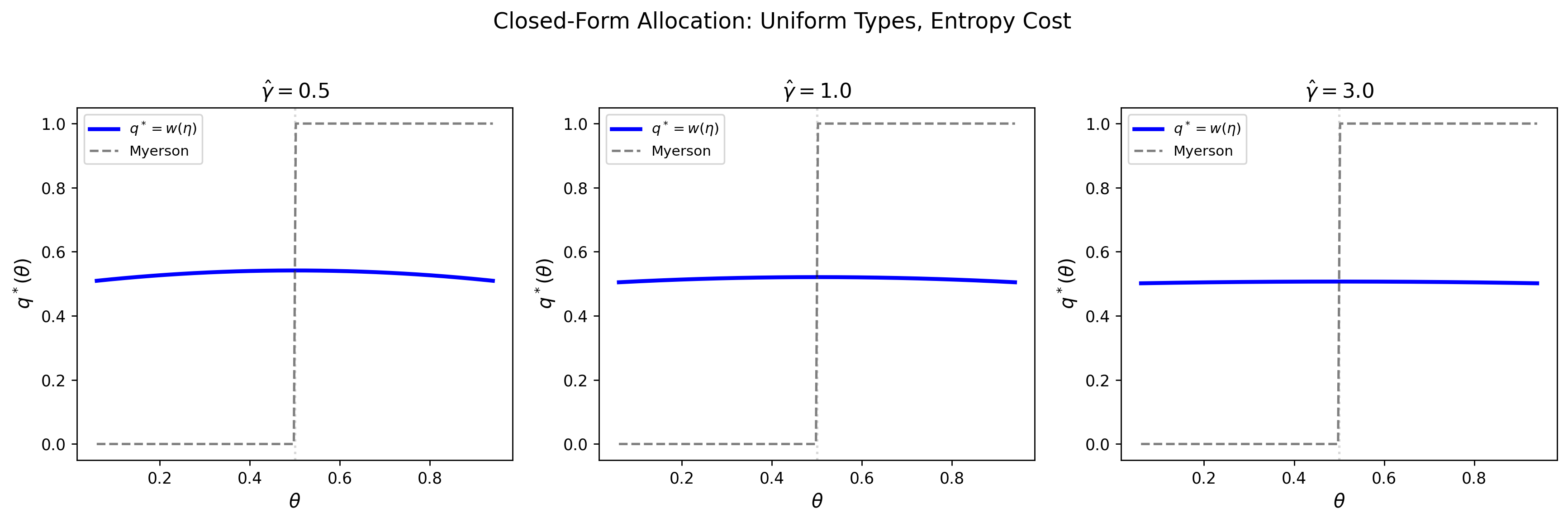}
\caption{Allocation $q^*(\theta) = w(\eta(\theta))$ for uniform types with entropy cost and Myerson multiplier $\lambda = 1-\theta$, at three values of $\hat\gamma$. Dashed: Myerson benchmark. With this multiplier, $\eta$ is uniformly small and $q^* \approx 1/2$ for all types. The optimal multiplier $\lambda^*$ concentrates investigation near $\theta_0$, as the numerical solution in Section~\ref{subsec:outer} shows.}
\label{fig:closedform}
\end{figure}

The variational problem~\eqref{eq:revenue_variational} determines the optimal multiplier $\lambda^*$. One natural candidate is the Myerson multiplier $\lambda(\theta) = 1-F(\theta)$, which sets the IC slack to zero at every type. For this candidate, the investigation intensity parameter on uniform types is $\eta(\theta) = \theta(1-\theta)/(2\hat\gamma) \leq 1/(8\hat\gamma)$, which is uniformly small. The allocation $q^*(\theta) = w(\eta(\theta)) \approx 1/2$ for all $\theta$, as Figure~\ref{fig:closedform} illustrates: every type is investigated, and the allocation is nearly uniform across the type space. The Hessian of~\eqref{eq:revenue_variational} at the Myerson multiplier has positive eigenvalues, confirming that Myerson is a saddle point rather than a maximum of the revenue functional.

\begin{remark}[The role of limited liability in the investigation benefit]\label{rem:benefit}
The investigation benefit arises from limited liability, which creates a transfer margin that signal-contingent pricing can exploit. Without limited liability, $d=1$ and the optimal experiment is binary by Corollary~\ref{cor:LL}; the transfer is pinned by the envelope and the investigation has no value beyond the allocation margin. With limited liability, $d=2$ and the seller gains a second instrument: she can charge $\bar p$ when the signal indicates a deviator and refund when it indicates a truthful reporter. This signal-contingent pricing generates revenue beyond what the envelope-determined transfer achieves. The seller always weakly benefits from having the investigation technology, since the no-investigation mechanism is a feasible choice.
\end{remark}

The optimal multiplier $\lambda^*$ departs from Myerson in the investigation region: non-Myerson multipliers concentrate investigation near $\theta_0$, matching the pattern in Section~\ref{subsec:outer}. The analytical characterization of $\lambda^*$ is a constrained variational problem with a free boundary at $\partial\mathcal{I}$, where $\lambda$ transitions between the interior regime and the corner $\lambda = 0$. The defining equation is transcendental, mixing $\lambda$ polynomially with $e^{\lambda\theta/(2\hat\gamma)}$ through the concavification identity $b/a = e^\eta$, so no closed-form solution exists. Section~\ref{subsec:outer} solves it numerically.

For any given multiplier $\lambda$, the seller's revenue under the investigation mechanism is
\begin{equation}\label{eq:R_invest}
    R_{\text{invest}}[\lambda] = \int_0^1 \bigl[\Phi(\theta)\, w\bigl(\eta(\theta;\lambda)\bigr) - 2\hat\gamma\, k\bigl(\eta(\theta;\lambda)\bigr)\bigr] f(\theta)\, d\theta,
\end{equation}
where $\eta(\theta;\lambda) = \lambda(\theta)\theta/(2\hat\gamma)$. Since the seller can always replicate the Myerson mechanism by choosing $F_\theta = \delta_1$ at every report, $\max_\lambda R_{\text{invest}}[\lambda] \geq R_{\text{Myerson}}$: the seller weakly benefits from having the investigation technology available.

%% =====================================================================
%% SECTION 7: COMPARATIVE STATICS
%% =====================================================================
\section{Comparative Statics and Welfare}\label{sec:compstats}

The structure of the optimal mechanism varies systematically with the primitives. This section characterizes how investigation intensity responds to the type, the cost parameter, and privacy restrictions, and derives the welfare implications of information acquisition.

Define the \textit{information intensity} at type $\theta$ as $I(\theta) = |\supp(F_\theta^*)| - 1 \in \{0, 1, 2\}$ and the \textit{investigation cost} as $K(\theta) = \int \psi(z) \dif F_\theta^*(z)$.

\begin{proposition}[Comparative statics]\label{thm:compstats}
Under Assumptions~\ref{ass:types}--\ref{ass:cost}:
\begin{enumerate}[label=(\alph*)]
    \item \textbf{Investigation concentrates near indifference.} $I(\theta)$ and $K(\theta)$ are maximized in a neighborhood of $\theta_0$ where $\Phi(\theta_0) = 0$. As $|\Phi(\theta)|$ grows, $I(\theta)$ decreases from $2$ to $1$ to $0$. The ternary region $\{\theta : I(\theta) = 2\}$ is connected and contains $\theta_0$.
    
    \item \textbf{Cost scaling.} Replacing $\psi$ with $\alpha\psi$ ($\alpha > 1$) shrinks the ternary region toward $\theta_0$ but does not change the maximum support size.
    
    \item \textbf{Privacy restrictions.} A per-report cost cap $\int \psi(z) \dif F_r(z) \geq -\bar{\psi}$ compresses the experiment's support points toward $z=1$. Welfare improves for types far from $\theta_0$, where the cap is slack, and may decrease near $\theta_0$, where the cap binds.
\end{enumerate}
\end{proposition}

The comparative statics trace how the investigation structure responds to the economic environment. Near the virtual-surplus zero-crossing $\theta_0$, both cutoffs $\kappa_x$ and $\kappa_p$ are close to $z=1$, creating the widest non-concave region in $M$ and the richest experiment. As the type moves away from $\theta_0$, the preferred action becomes clear without investigation: one cutoff passes $z=1$, the experiment transitions from ternary to binary to degenerate, and investigation intensity falls monotonically. The investigation region is thus a band centered at the zero-crossing, with the ternary sub-region as its innermost core.

Scaling the cost by a factor $\alpha > 1$ increases the curvature of $\psi$ without moving the kink locations, which are determined by the dual value $P$. The non-concave regions of $M$ shrink because the cost curvature makes $M$ ``more concave'' on each piece, but the maximum support size remains $d+1$ because the number of kinks is unchanged. The effect is a narrowing of the investigation band, not a change in its structure.

A privacy restriction that caps the per-report information expenditure compresses the experiment's support points toward $z=1$. For types far from $\theta_0$, the cap is slack and the constrained optimum coincides with the unconstrained one. For types near $\theta_0$, the cap binds: the seller acquires less information than she would like, and the welfare effect is ambiguous because the cap reduces both the investigation benefit and the investigation cost. The proof is in Appendix~\ref{app:compstats}.

\begin{figure}[htbp]
\centering
\includegraphics[width=0.95\textwidth]{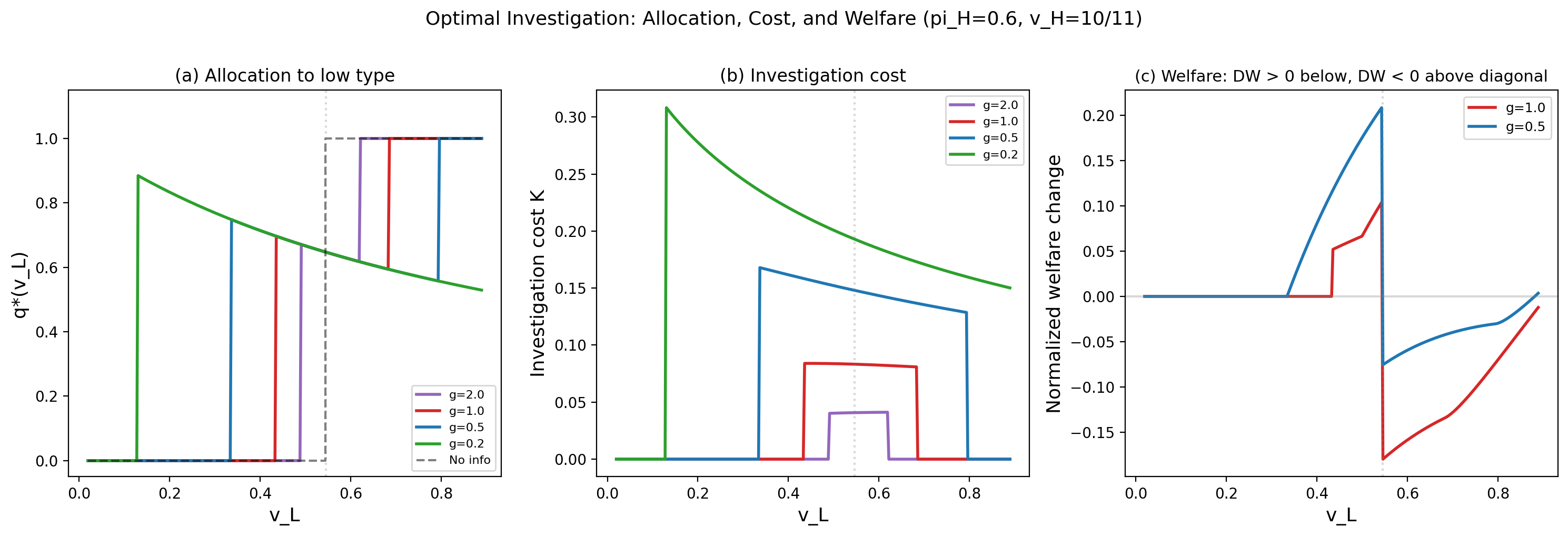}
\caption{Investigation, allocation, and welfare for $\pi_H = 0.6$, $v_H = 10/11$, at four values of $\gamma$. Panel~a: the allocation $q^*(v_L)$ is smoothed relative to the step-function Myerson benchmark, shown dashed. Panel~b: investigation cost peaks near the indifference boundary $v_L = \pi_H v_H$ and widens with cheaper information. Panel~c: welfare improves below the diagonal, $\bar\Delta W > 0$, and worsens above, $\bar\Delta W < 0$.}
\label{fig:welfare}
\end{figure}

The investigation region narrows as information becomes expensive (Figure~\ref{fig:welfare}).

\begin{proposition}[Investigation region]\label{prop:region}
Fix $\pi_H$ and the entropy cost $\psi(z) = -2\gamma\log z$. The investigation region $\mathcal{I}(\gamma) = \{v_L : F^*_{v_L} \neq \delta_1\}$ satisfies:
\begin{enumerate}[label=(\alph*)]
    \item $\mathcal{I}(\gamma)$ is an interval centered approximately at the indifference boundary $v_L = \pi_H v_H$.
    \item The width $|\mathcal{I}(\gamma)|$ is decreasing in $\gamma$: $|\mathcal{I}(\gamma)| \propto \gamma^{-1}$ for large $\gamma$.
    \item Inside $\mathcal{I}(\gamma)$, the allocation probability $q^*(v_L) \in (0,1)$, strictly between the Myerson extremes. Information acquisition smooths the step function.
    \item As $\gamma \to 0$ (cheap information), $\mathcal{I} \to (0, v_H)$: the seller investigates everywhere.
    \item As $\gamma \to \infty$ (expensive information), $\mathcal{I} \to \emptyset$: no investigation.
\end{enumerate}
\end{proposition}

The investigation region is an interval because the cutoffs $\kappa_x$ and $\kappa_p$ vary continuously in $v_L$, so the condition $a_{xp} < 1 < b_{xp}$ that triggers investigation holds on a connected set. The width of this interval scales as $1/\gamma$: by Proposition~\ref{prop:closedform}, the concavification endpoints $a$ and $b$ converge to the kink $\kappa$ at rate $1/\gamma$, so the range of types where $z=1$ falls between $a$ and $b$ shrinks proportionally.

Inside the investigation region, the allocation probability is strictly interior: $q^* = w(\eta) \in (0,1)$. This is the hallmark of information acquisition in screening. The Myerson mechanism assigns $q \in \{0,1\}$ deterministically; investigation replaces this sharp boundary with a gradual transition governed by the universal function $w$. Types near the zero-crossing receive an allocation probability close to $1/2$, while types at the edge of the investigation region receive allocations close to $0$ or $1$.

The limiting behavior is intuitive. When information is cheap, $\gamma \to 0$, the concavification endpoints spread: $a \to 0$ and $b \to \infty$, so the condition $a < 1 < b$ holds for every type with positive virtual surplus, and the seller investigates the entire type space. When information is expensive, $\gamma \to \infty$, the endpoints converge: $a$ and $b$ both approach $\kappa$, the non-concave region collapses, and no type justifies the cost of investigation.

\begin{proposition}[Welfare]\label{prop:welfare}
Fix $\pi_H$ and the entropy cost. Let $W^*$ denote total welfare under the optimal mechanism and $W^{NI}$ under the no-information mechanism.
\begin{enumerate}[label=(\alph*)]
    \item \textbf{Seller always benefits:} $U^*_S \geq U^{NI}_S$ for all parameter values. The seller can always replicate the no-information mechanism by choosing the degenerate experiment.
    \item \textbf{Below the indifference diagonal} ($v_L < \pi_H v_H$): $W^* \geq W^{NI}$. The no-information mechanism inefficiently excludes the low type. Investigation enables the seller to serve some low types (efficiency gain) while extracting surplus through screening (rent extraction). The efficiency gain dominates.
    \item \textbf{Above the indifference diagonal} ($v_L > \pi_H v_H$): $W^* \leq W^{NI}$. The no-information mechanism is already efficient (both types trade at price $v_L$). Investigation introduces screening distortions that reduce welfare.
\end{enumerate}
\end{proposition}

The welfare result reveals how investigation interacts with market efficiency. The seller always weakly benefits, since the degenerate experiment $F = \delta_1$ is a feasible choice that replicates the no-information mechanism; any departure from it is voluntary and revenue-improving.

The effect on total welfare depends on whether the no-information mechanism is already efficient. Below the indifference diagonal, $v_L < \pi_H v_H$, the no-information mechanism posts $p = v_H$ and excludes the low type entirely, leaving social surplus $\pi_L v_L$ on the table. Investigation enables the seller to serve some low types at a screened price, capturing part of this surplus as revenue while generating an efficiency gain. The net welfare change is $\Delta W = \pi_L v_L q^* - K$, which is nonnegative: the seller's revealed preference ensures $\pi_L \bar p^* \geq K$, and $\bar p^* \leq v_L q^*$ by IR, so the social surplus from trade exceeds the investigation cost.

Above the diagonal, $v_L > \pi_H v_H$, the no-information mechanism already achieves the first-best: both types trade at price $v_L$, and $q = 1$ for all. Investigation can only distort this efficient allocation by excluding the low type on some signal realizations, reducing total surplus. The investigation cost compounds the loss. Welfare decreases whenever the seller investigates above the diagonal.

%% =====================================================================
%% SECTION 8: IMPLEMENTATION
%% =====================================================================
\section{Implementation}\label{sec:implementation}

The abstract three-region partition translates into a concrete menu that the seller can offer. The mechanism takes the form of a tiered program in which each buyer self-selects into one of three service levels, each corresponding to a different level of investigation. The structure closely resembles existing institutions: loyalty programs in retail, staged underwriting in insurance, and tiered verification in digital pricing.

\paragraph{Tier~1, Region~I: no investigation.} High-type buyers purchase the good outright at a posted price $p_H$. The seller acquires no information.

\paragraph{Tier~2, Region~II: binary test.} Intermediate-type buyers pay a base fee and undergo a binary test such as a credit check. Two outcomes: \textit{pass}, the buyer receives the good at a discount and the fee is refunded; \textit{fail}, the fee is retained and the good is withheld.

\paragraph{Tier~3, Region~III: ternary test.} Near-indifference types enter a three-outcome verification program. Outcome~1, strongly consistent with the truthful report: good at maximum discount, fee refunded. Outcome~2, ambiguous evidence: intermediate treatment, good at full price or refund, depending on which margin is active. Outcome~3, evidence of misreport: no good, no refund.

This tiered structure maps directly onto existing market institutions. Purchasing the good outright corresponds to opting out of a loyalty program; the binary test corresponds to a standard credit check or eligibility screen; and the ternary test corresponds to a multi-stage underwriting process with an intermediate referral category.

%% =====================================================================
%% SECTION 9: DISCUSSION
%% =====================================================================
\section{Discussion}\label{sec:discussion}

This section draws out the broader implications of the results. We discuss testable predictions for consumer lending and digital pricing, explain why the ternary bound arises from two margins rather than one or three, and position the paper relative to the information design and mechanism design literatures.

\subsection{Testable implications}

The model generates three predictions amenable to empirical testing. First, \textit{investigation intensity should peak near the approval threshold}. In consumer lending, investigation effort should be highest for applicants near the bank's indifference margin, not for the riskiest or safest applicants. Second, \textit{investigation should produce coarse classifications}. The ternary bound predicts that optimal data acquisition partitions customers into at most three segments, matching the coarse tier structure of loyalty programs and the pass/refer/fail classification in automated underwriting systems. Third, \textit{the investigation region should narrow as data costs rise}. Tighter privacy regulation (increasing the effective $\gamma$) should shrink the band of types subject to investigation while leaving the three-region structure intact.

\subsection{Why two margins, not one or three}

The ternary bound reflects two frictions acting simultaneously. The \textit{allocation margin} exists because the seller screens: she excludes low-value buyers to extract surplus. The \textit{transfer margin} exists because of limited liability: $p \in [0, \pbar]$ makes the transfer a constrained variable, not a residual. Without limited liability, $d=1$ and signals are binary (Section~\ref{subsec:benchmark}). With limited liability, $\kappa_p \neq \kappa_x$ generically, creating the second kink.

The effective policy dimension counts \textit{decisions that the signal can improve}. With $d$ margins, at most $d$ actions change, requiring at most $d+1$ likelihood-ratio regions. More complex signals are suboptimal: the convex cost penalizes dispersion beyond $d+1$ points. The bound is tight because the cost of additional complexity exceeds its benefit once all margins are resolved.

\subsection{The design-side logistic: connection to rational inattention}

The universal function $w(\eta)$ invites comparison with the logistic function $\ell(\eta) = 1/(1+e^{-\eta})$ from rational inattention \citep{matejka2015}. Both arise from entropy-cost optimization with binary actions, both satisfy $w(0) = \ell(0) = 1/2$ and $w(\infty) = \ell(\infty) = 1$, and both are strictly increasing. But they are not the same function, and the difference has economic content.

In rational inattention, a decision-maker \textit{processes} a signal about an unknown state and chooses an action. The entropy cost penalizes deviation from the prior. The optimal action probability is the logistic $\ell(\eta)$, where $\eta$ is the payoff difference scaled by the cost parameter. In our model, the seller \textit{designs} a signal subject to Bayes plausibility, the mean-one constraint $\int z\dif F = 1$. The optimal allocation is $w(\eta)$, where $\eta$ is the investigation intensity.

The key result is that $w(\eta) < \ell(\eta)$ for all $\eta > 0$: the design-side allocation is strictly below the processing-side allocation at every investigation intensity. The gap peaks at approximately $\eta \approx 1.8$, where the designer allocates about 10\% less than the RI agent. The source of the gap is the Bayes-plausibility constraint: the designer must ``budget'' her signal so that the mean likelihood ratio equals one, which forces a more conservative allocation than entropy maximization alone would produce. The function $w$ is thus the design-side analog of the logistic, the allocation rule that emerges when the principal chooses \textit{what to learn} rather than how to act on \textit{what she already knows}.

\subsection{Related work}

In \textit{information design}, \citet{kamenica2011} establish concavification; \citet{dworczak2019} and \citet{dworczak2024} develop duality methods; \citet{kleiner2021} characterize extreme points under majorization. Our setting adds a strictly convex cost $\psi$, which drives the support bound absent in costless persuasion. \citet{lyu2024} impose exogenous cardinality constraints; we derive coarseness endogenously. \citet{kolotilin2025} unite persuasion and delegation; \citet{candogan2023} study disclosure to privately informed agents; \citet{bardhi2024} studies strategic attribute sampling.\footnote{In \citet{bardhi2024}, the agent samples attributes to influence the principal's decision; information is free but limited in quantity. In our setting, the principal acquires information at a convex cost to verify the agent's report; the signal space is unrestricted but costly.}

In \textit{mechanism design with information}, \citet{bergemann2015} bound welfare across information structures; \citet{roesler2017} study buyer-optimal learning; \citet{frick2026} study precise exogenous seller information; \citet{heumann2020} studies sequential screening with ex post participation; \citet{ditillio2021} study strategic sample selection. In \textit{rational inattention}, \citet{matejka2015} show that entropy-cost optimization with discrete actions produces logistic choice probabilities. Our universal function $w(\eta)$ is the design-side counterpart: it arises from the same entropy cost but under Bayes plausibility rather than a prior constraint, and it is strictly below the logistic.

In \textit{costly verification}, \citet{townsend1979} and \citet{benporath2014} study binary verification; \citet{erlanson2020} address collective decisions. Our framework endogenizes verification precision, connecting verification to information design. The closest precursor is \citet{georgiadis2020}: our effective-policy-dimension theorem generalizes their binary-signal result, $d=1$, to arbitrary $d$.

%% =====================================================================
%% SECTION 10: CONCLUSION
%% =====================================================================
\section{Conclusion}\label{sec:conclusion}

Two institutional features of screening shape the complexity of optimal information. Limited liability determines how many signal outcomes the seller needs: without it, the dimension drops to one and signals are binary; with it, the dimension rises to two and signals are ternary. The cost of information determines where the seller investigates: in a band around the exclusion threshold, whose width scales as $1/\gamma$.

Together, these forces dissolve the Myerson exclusion boundary. The discontinuous step function $q \in \{0,1\}$, the most recognized feature of optimal auction design, is an artifact of the seller's inability to investigate. With investigation, the sharp cutoff is replaced by a smooth transition governed by a universal function $w(\eta)$. Every marginal buyer has a strictly positive probability of trade. The function $w$ is the design-side analog of the logistic function from rational inattention: both arise from entropy-cost optimization with binary actions, but $w(\eta) < \ell(\eta)$ for all $\eta > 0$ because the signal designer faces a Bayes-plausibility constraint that the rational-inattention agent does not. The gap quantifies the cost of designing information rather than merely processing it.

For the entropy cost, the concavification admits a closed-form solution through the identity $b/a = e^\eta$, yielding a universal allocation function $w(\eta)$ that governs the smoothing of the Myerson step function whenever investigation is active. The optimal multiplier $\lambda^*$ that determines where and how intensely the seller investigates satisfies a transcendental equation that precludes closed-form solution; the numerical saddle-point computation confirms that $\lambda^*$ departs from the Myerson multiplier and concentrates investigation near the virtual-surplus zero-crossing.

The normative implication is direct: privacy regulators can cap signal complexity at $d+1$ outcomes without reducing efficiency. In the canonical screening environment, three outcomes suffice; only richer action spaces require more.

The framework extends to binding participation constraints with $d$ unchanged, Appendix~\ref{app:IRH}; to mutual-information costs with an additional atom at zero, Remark~\ref{rem:MI}; to quality choice with $d=3$ and three distinct thresholds confirmed numerically, Example~\ref{ex:quality}; and to the full saddle-point computation, Section~\ref{subsec:outer}, where the transfer margin departs from the Myerson benchmark and investigation concentrates at the virtual-surplus zero-crossing.

%% BIBLIOGRAPHY
%% =====================================================================
\newpage
%\bibliographystyle{ecta}

%% =====================================================================
%% APPENDIX
%% =====================================================================
\newpage
\appendix
\section{Appendix: Proofs}\label{app:proofs}

\subsection{Proof of Proposition~\ref{prop:localIC}}

This is standard; we include it for completeness. Under single crossing, where $\theta x - p$ is supermodular in $(\theta, x)$, incentive compatibility requires the allocation schedule $q(\theta)$ to be nondecreasing. The envelope theorem for arbitrary choice sets \citep{milgrom2002} then delivers $U(\theta) = U(\thetaubar) + \int_{\thetaubar}^\theta q(s)\dif s$. Participation at the bottom, $U(\thetaubar) \geq 0$, completes the characterization. See \citet[Chapter~2]{laffont2002}. \hfill$\square$

\subsection{Proof of Lemma~\ref{lem:binary}}\label{app:binary}

\textit{Step 1: Lagrangian separability.} After dualizing the $N-1$ local downward IC constraints with multipliers $\lambda_1, \ldots, \lambda_{N-1}$ and the bottom-type participation constraint with multiplier $\mu$, the Lagrangian is
\[
    L = \sum_{j=1}^{N}\Bigl[\int_0^\infty M_{j,\Lambda}(z)\dif F_j(z)\Bigr] + \text{(terms not involving $F$)}.
\]
Each $F_j$ appears in exactly one summand, so the experiments are separable given $\Lambda$.

\textit{Step 2: Binary state.} At report $\theta_j$, the IC$_{j,j+1}$ constraint involves the deviation payoff $\int(\theta_{j+1}x_j(z) - p_j(z))z\dif F_j(z)$, where the $z$-weighting arises from the likelihood ratio of type $\theta_{j+1}$'s signal distribution relative to type $\theta_j$'s. The truthful payoff integrates against $\dif F_j(z)$. Thus $z$ is the likelihood ratio for the binary hypothesis $\{\theta_j, \theta_{j+1}\}$.

\textit{Step 3: Sufficiency.} Under monotone $q$, binding local IC implies global IC (Proposition~\ref{prop:localIC}). Therefore, no information about non-adjacent types is needed: the seller's problem at report $\theta_j$ is a binary-state information-design problem over $z$ with $\int z\dif F_j = 1$. \hfill$\square$

\subsection{Proof of Lemma~\ref{lem:strongdual}}\label{app:strongdual}

For fixed experiments $\{F_j\}$, the screening problem~\eqref{eq:lagrangian} is a finite-dimensional linear program in the allocation and transfer variables $(x_j(z), p_j(z))$, parameterized by $\Lambda$. By the Lagrange multiplier theorem \citep{luenberger1969}, strong duality holds and the infimum over $\Lambda \geq 0$ is attained. Convexity of $L$ in $\Lambda$ as a supremum of affine functions, is immediate. \hfill$\square$

\subsection{Proof of Lemma~\ref{lem:nokink}}\label{app:nokink}

At any kink $\kappa$ of $P_{j,\Lambda}$, the slope of $P$ increases since $P$ is convex. Hence $M = P - \psi$ has a strict convex kink at $\kappa$: the left derivative $M'(\kappa^-)$ is strictly less than the right derivative $M'(\kappa^+)$. This means $M$ is not concave at $\kappa$, so $M_{j,\Lambda}(\kappa) < M^c_{j,\Lambda}(\kappa)$. Any experiment placing mass $w > 0$ at $\kappa$ satisfies $\int M \dif F \leq \int M^c \dif F - w(M^c(\kappa) - M(\kappa)) < M^c(1)$, where the last inequality uses Jensen applied to the concave $M^c$ and the mean-one constraint $\int z\dif F = 1$. Since $M^c(1)$ is attainable, no optimal experiment places mass at $\kappa$. \hfill$\square$

\subsection{Proof of Theorem~\ref{thm:d2}}\label{app:thm_d2}

The proof has three parts: the threshold structure, the kink count, and the support bound.

\textit{Part~1: Threshold structure.} At report $\theta_j$ with multiplier vector $\Lambda$, the dual post-signal objective is
\[
    P_{j,\Lambda}(z) = \max_{x\in[0,1],\, p\in[0,\pbar]} \bigl\{ \underbrace{(\lambda_{j-1}\theta_j - \lambda_j\theta_{j+1}z)}_{A_j(z)}\, x + \underbrace{(\pi_j - \lambda_{j-1} + \lambda_j z)}_{B_j(z)}\, p \bigr\},
\]
where $\lambda_0 := \mu$ (the IR multiplier) and $\lambda_j$ is the IC$_{j,j+1}$ multiplier. Because $A_j$ is affine and decreasing in $z$, and $B_j$ is affine and increasing in $z$, and because $(x,p)$ are binary controls, the optimizer is threshold-based: $x^*(z) = \mathbf{1}[z \leq \kappa_x]$ and $p^*(z) = \pbar\,\mathbf{1}[z > \kappa_p]$, where
\[
    \kappa_x(j,\Lambda) = \frac{\lambda_{j-1}\theta_j}{\lambda_j\theta_{j+1}}, \qquad \kappa_p(j,\Lambda) = \frac{\lambda_{j-1} - \pi_j}{\lambda_j}.
\]

\textit{Part~2: Kink count.} $P_{j,\Lambda}$ is the maximum of four affine functions corresponding to the four combinations of $x \in \{0,1\}$ and $p \in \{0,\pbar\}$. The active piece switches at $\kappa_x$ and at $\kappa_p$, producing exactly two kinks when $\kappa_x \neq \kappa_p$.\footnote{The condition $\kappa_x = \kappa_p$ is equivalent to $\lambda_{j-1}(1-\theta_j/\theta_{j+1}) = \pi_j$, a single equation in multiple free parameters. It is therefore codimension-1 and fails generically.}

\textit{Part~3: Support bound.} Define $M_{j,\Lambda}(z) = P_{j,\Lambda}(z) - \psi(z)$. Since $P$ is piecewise affine and convex with at most 2 kinks and $\psi$ is strictly convex by Assumption~\ref{ass:cost}, $M$ is piecewise strictly concave with at most 2 kinks. Its concave envelope $M^c$ at $z=1$ is achieved by a measure supported on points where $M = M^c$. The concavification of a piecewise strictly concave function with $k$ kinks on a connected non-concave region containing $z=1$ is supported on at most $k+1$ points. Each non-concave piece contributes at most one concavification endpoint; by strict concavity, $M < M^c$ strictly in the interior of each non-concave piece, so no interior mass is optimal. With $k \leq 2$ kinks, $|\supp(F_j^*)| \leq 3$.

That \textit{every} optimum satisfies this bound, not just some, follows from strict concavity: if $F^*$ placed positive mass at any point $z_0$ where $M(z_0) < M^c(z_0)$, then replacing the mass at $z_0$ with mass at the concavification endpoints while preserving the mean would strictly increase $\int M \dif F$, contradicting optimality. \hfill$\square$

\subsection{Proof of Lemma~\ref{lem:minimax}}\label{app:minimax}

We verify the hypotheses of Sion's minimax theorem \citep{sion1958}.

\textit{Step~1: Domain restrictions.} For any finite revenue bound $\bar{\Pi}$, the cost constraint $\int\psi(z)\dif F_j(z) \leq \bar{\Pi}/\pi_j$ and Assumption~\ref{ass:cost}, namely $\psi(z)\to+\infty$ as $z\to 0^+$ and $\psi$ strictly convex, imply that $F_j$ is tight: for any $\varepsilon > 0$, there exist $0 < \varepsilon' < K' < \infty$ such that $F_j([\varepsilon',K']) \geq 1 - \varepsilon$. Restricting to $F_j \in \mathcal{F}_{[\varepsilon,K]}$ , the set of mean-one distributions on $[\varepsilon,K]$, for sufficiently small $\varepsilon$ and large $K$ is without loss.

\textit{Step~2: Convexity-concavity.} The Lagrangian $L(\Lambda, F) = \sum_j \int M_{j,\Lambda}(z)\dif F_j(z) + \text{terms not involving }F$ is convex in $\Lambda$, as a supremum of affine functions of $\Lambda$, and linear in $F$, hence both convex and concave.

\textit{Step~3: Compactness and semicontinuity.} $\Lambda$ ranges over $[0,\bar\lambda]^{N-1}$, which is compact. $\mathcal{F}_{[\varepsilon,K]}$ is compact in the weak topology by Prohorov's theorem. $L$ is continuous in $F$ since the integrand is bounded and continuous on $[\varepsilon,K]$ and continuous in $\Lambda$ because $M_{j,\Lambda}(z)$ is continuous in $\Lambda$ for each $z$ and Lebesgue's dominated convergence applies.

\textit{Step~4: Sion's theorem.} The hypotheses of convex-concavity, compact domains, and appropriate semicontinuity are satisfied. Hence $\sup_F \inf_\Lambda L = \inf_\Lambda \sup_F L$ and a saddle point $(\Lambda^*, F^*)$ exists. \hfill$\square$

\subsection{Proof of Theorem~\ref{thm:fullchar}}\label{app:fullchar}

\textit{Part~(a): No investigation at the top.} At the top type $\theta_N$, there is no upward deviator. The IC multiplier $\lambda_N$ is zero. The Lagrangian contribution from report $\theta_N$ involves only the IR term, which is maximized by a posted price. No experiment is needed: $F_N^* = \delta_1$.

\textit{Part~(b): Ternary bound.} Immediate from Theorem~\ref{thm:d2} applied at each report.

\textit{Part~(c): Monotone allocation.} The allocation $q(\theta_j) = \int x(j,z)\dif F_j^*(z)$ may not be monotone in $j$ at the saddle point. When monotonicity fails, we apply the standard ironing procedure \citep{myerson1981}: pool adjacent types with non-monotone $q$ values and replace their allocations with the common mean. The ironed allocation satisfies monotonicity and has weakly higher revenue by convexity of the objective in $q$. At each ironed report, the local binary-verification structure is preserved since the multipliers adjust but the pointwise problem retains $d=2$, so the ternary bound survives ironing.

\textit{Part~(d): Envelope.} With IC$_{j,j+1}$ binding at the saddle point, $U(\theta_{j+1}) - U(\theta_j) = (\theta_{j+1}-\theta_j)q(\theta_j)$. Telescoping from $U(\theta_1) = 0$ (IR binding) gives the stated formula.

\textit{Part~(e): Three-region partition.} At the saddle point $\Lambda^*$, the function $M_{j,\Lambda^*}$ has cutoffs $\kappa_x(j)$ and $\kappa_p(j)$ that vary with $j$. (i)~When both cutoffs are far from $z=1$ (equivalently, $M_{j,\Lambda^*}$ is concave at $z=1$), the experiment is degenerate. (ii)~When one non-concave region contains $z=1$, the experiment is binary. (iii)~When the merged non-concave region $Z_{xp}$ contains $z=1$, the experiment is ternary. The ternary region is connected because the cutoffs vary continuously in $j$.

\textit{Part~(f): IR-H binding.} See Appendix~\ref{app:IRH} below.

\textit{Part~(g): Uniqueness.} Under strict complementary slackness, the saddle-point $\Lambda^*$ is unique because the Lagrangian is strictly convex in $\Lambda$ at the optimum. Given $\Lambda^*$, the experiment $F_j^*$ is unique at each report where the concavification is non-degenerate, since strict concavity of $M$ on each piece implies the concavification weights are uniquely determined. \hfill$\square$

\subsection{The case when IR-H binds}\label{app:IRH}

When the highest type's participation constraint binds, the dual acquires an additional multiplier $\mu_N \geq 0$ for IR at $\theta_N$. The Lagrangian at report $\theta_j$ becomes:
\[
    L_j = \int \bigl[P_{j,\Lambda,\mu}(z) - \psi(z)\bigr]\dif F_j(z),
\]
where $P_{j,\Lambda,\mu}$ now includes terms from $\mu_N$. Since $P$ remains piecewise affine in $z$ with two controls $(x,p)$, the effective policy dimension stays at $d=2$. The cutoff formulas change but the qualitative structure, two thresholds, piecewise strictly concave $M$, support $\leq 3$, is preserved. The three-region partition persists with shifted boundaries. \hfill$\square$

\subsection{Proof of Theorem~\ref{thm:EPD}}\label{app:EPD}

\textit{Part~(i): Piecewise strict concavity.} By hypothesis, $P_\Lambda(z) = \sup_{a\in A}\{\alpha_\Lambda(a) + z\beta_\Lambda(a)\}$ is piecewise affine and convex with at most $d$ kinks. Subtracting the strictly convex $\psi$ yields $M_\Lambda = P_\Lambda - \psi$, which is strictly concave on each affine piece because $-\psi$ is strictly concave. The kinks of $M_\Lambda$ coincide with those of $P_\Lambda$.

\textit{Part~(ii): Support bound.} Consider the concavification of $M_\Lambda$ at $z=1$, namely $M_\Lambda^c(1) = \sup\{\int M_\Lambda\dif F : \int z\dif F = 1\}$. If $M_\Lambda(1) = M_\Lambda^c(1)$, the optimum is $F^* = \delta_1$ with support 1. Otherwise, $z=1$ lies in a non-concave region of $M_\Lambda$. This region spans at most $d$ kinks, dividing it into at most $d+1$ strictly concave pieces. The concave envelope $M_\Lambda^c$ is affine between consecutive support points, and each strictly concave piece can contain at most one support point: if it contained two, strict concavity would imply $M < M^c$ between them, contradicting $M = M^c$ at support points. Hence $|\supp(F^*)| \leq d+1$.

That every optimum, not just some, satisfies the bound follows from the strict Jensen argument: positive mass at any $z_0$ with $M_\Lambda(z_0) < M_\Lambda^c(z_0)$ yields a strictly lower objective, contradicting optimality.

\textit{Part~(iii): Tightness.} For generic $\Lambda$, all $d$ kinks are distinct and lie in a non-concave region containing $z=1$. In this case the concavification places positive weight on exactly $d+1$ points, one in each of the $d+1$ concave pieces. This is achieved by choosing parameters such that $z=1$ lies in the joint non-concave region spanning all kinks.

\textit{Part~(iv): Minimax survival.} At the saddle point $(\Lambda^*, F^*)$ from Lemma~\ref{lem:minimax}, $F^*$ maximizes $\int M_{\Lambda^*}\dif F$ subject to $\int z\dif F = 1$. This is precisely the information-design problem for $M_{\Lambda^*}$, to which parts~(i)--(ii) apply directly. Hence $|\supp(F^*)| \leq d+1$. \hfill$\square$

\subsection{Proof of Proposition~\ref{thm:convergence}}\label{app:convergence}

\textit{Step~1: Compactness.} For each $N$ and report $\theta_j^N$, the cost bound $\int\psi(z)\dif F_j^N(z) \leq \bar\Pi / \pi_j^N$ , where $\bar\Pi$ is the upper bound on total revenue independent of $N$, and Assumption~\ref{ass:cost}, which gives $\psi(z)\to\infty$ as $z\to 0^+$, prevent mass from accumulating near zero. The mean-one constraint $\int z\dif F_j^N = 1$ prevents mass from escaping to infinity by Markov's inequality since $F_j^N([0,K]) \geq 1 - 1/K$. Hence each $F_j^N$ lies in a tight, uniformly bounded family of probability measures on $[\varepsilon,K]$ for some $\varepsilon > 0$, $K < \infty$ independent of $N$.

\textit{Step~2: Subsequence extraction.} The allocation schedule $q^N(\cdot)$ is a nondecreasing function on a grid approximating $\Theta$. By Helly's selection theorem, there is a subsequence $N_k$ such that $q^{N_k}(\theta) \to q^*(\theta)$ at every continuity point of $q^*$, and $q^*$ is nondecreasing. For each $\theta \in \Theta$, the experiment $F_\theta^{N_k}$ (defined by interpolation or nearest-grid assignment) converges weakly to some $F_\theta^*$ by Prohorov's theorem applied to the tight family.

\textit{Step~3: IC passage.} The discrete envelope $U^N(\theta_j) = \sum_{k<j}(\theta_{k+1}-\theta_k)q^N(\theta_k)$ converges, by uniform convergence of the Riemann sums, to $U^*(\theta) = \int_{\thetaubar}^\theta q^*(s)\dif s$. This is the continuum envelope formula. Local IC at adjacent grid types converges to the differential IC condition $U'(\theta) = q(\theta)$ a.e. Global IC follows from monotonicity of $q^*$ and the envelope representation.

\textit{Step~4: Support preservation.} Each $F_j^N$ is supported on at most 3 points in the compact set $[\varepsilon,K]$. Write $F_j^N = \sum_{i=1}^{3} w_i^N \delta_{z_i^N}$ where $w_i^N = 0$ when the support has fewer points. By compactness of $[\varepsilon,K]^3 \times \Delta_3$, pass to a further subsequence where $(z_1^{N_k},z_2^{N_k},z_3^{N_k}) \to (z_1^*,z_2^*,z_3^*)$ and $(w_1^{N_k},w_2^{N_k},w_3^{N_k}) \to (w_1^*,w_2^*,w_3^*)$. The weak limit is $F_j^* = \sum_i w_i^*\delta_{z_i^*}$, supported on at most 3 points since some $z_i^*$ may coincide, reducing the support size. Hence $|\supp(F_\theta^*)| \leq 3$ for a.e.\ $\theta$. \hfill$\square$

\subsection{Proof of Proposition~\ref{prop:closedform}}\label{app:closedform}

On the left piece $M_1(z) = \alpha_1 + \beta_1 z + 2\gamma\log z$ and the right piece $M_2(z) = \alpha_2 + \beta_2 z + 2\gamma\log z$, with $\Delta\beta = \beta_2 - \beta_1 > 0$ and $\alpha_1 - \alpha_2 = \Delta\beta\cdot\kappa$ from continuity at the kink. The tangency conditions $M_1'(a) = M_2'(b)$ yield $\beta_1 + 2\gamma/a = \beta_2 + 2\gamma/b$, hence $2\gamma(b-a)/(ab) = \Delta\beta$. The chord condition $M_1'(a) = [M_2(b)-M_1(a)]/(b-a)$ simplifies to $\Delta\beta\cdot\kappa = 2\gamma\log(b/a)$, giving $b/a = e^{\eta}$ with $\eta = \Delta\beta\cdot\kappa/(2\gamma)$. Substituting back: $b = 2\gamma(e^{\eta}-1)/\Delta\beta$ and $a = b/e^{\eta} = 2\gamma(1-e^{-\eta})/\Delta\beta$. The mean-one constraint $wa + (1-w)b = 1$ gives $w = (b-1)/(b-a)$. Using $2\gamma/\Delta\beta = \kappa/\eta$, the denominator is $b - a = (\kappa/\eta)(e^\eta + e^{-\eta} - 2)$ and the numerator is $b - 1 = (\kappa/\eta)(e^\eta - 1) - 1 = [\kappa(e^\eta - 1) - \eta]/\eta$. Hence $w = [\kappa(e^\eta-1)-\eta]/[\kappa(e^\eta+e^{-\eta}-2)]$. When $\kappa = 1$, this reduces to $(e^\eta - 1 - \eta)/(e^\eta + e^{-\eta} - 2)$. \hfill$\square$

\subsection{Proof of Proposition~\ref{thm:compstats}}\label{app:compstats}

\textit{Part~(a): Investigation concentrates near indifference.} The virtual surplus at type $\theta$ is $\Phi(\theta) = \theta - (1-F(\theta))/f(\theta)$. At the saddle point, the cutoffs satisfy
\[
    \kappa_x(\theta,\Lambda^*) = \frac{\lambda(\theta^-)v_L(\theta)}{\lambda(\theta^+)v_H(\theta)}, \qquad \kappa_p(\theta,\Lambda^*) = \frac{\lambda(\theta^-) - f(\theta)\Delta\theta}{\lambda(\theta^+)}.
\]
When $|\Phi(\theta)|$ is large, the Myerson allocation $q(\theta)$ is far from the boundary (either 0 or 1 with high certainty). The multiplier $\lambda(\theta)$ is approximately $1-F(\theta)$ under the Myerson benchmark, and the cutoffs $\kappa_x$, $\kappa_p$ are far from $z=1$. Hence $M_\theta$ is concave at $z=1$ and the experiment is degenerate. As $\theta$ approaches $\theta_0$ where $\Phi(\theta_0)=0$, the cutoffs approach $z=1$ from opposite sides, the non-concave region of $M_\theta$ begins to contain $z=1$, and the experiment transitions from degenerate to binary to ternary.

\textit{Part~(b): Cost scaling.} Replacing $\psi$ with $\alpha\psi$ scales $\gamma \to \alpha\gamma$ in the entropy-cost case. The concavification points $a_{xp}$ and $b_{xp}$ depend on $\gamma$: larger $\gamma$ widens the non-concave region, but also increases the cost of investigation, reducing the range of types where investigation is profitable. The ternary region, defined by $a_{xp}(\Lambda^*) < 1 < b_{xp}(\Lambda^*)$ with the additional condition that both non-concave sub-regions merge, shrinks toward $\theta_0$ as $\alpha$ increases.

\textit{Part~(c): Privacy restrictions.} A per-report cost cap $\int\psi(z)\dif F_r \leq \bar\psi$ constrains the experiment's support points to lie closer to $z=1$ (the degenerate experiment satisfies the cap trivially). For types far from $\theta_0$ where the cap is slack, the constrained and unconstrained optima coincide. For types near $\theta_0$ where the cap binds, the seller acquires less information, and welfare may decline relative to the unconstrained optimum. \hfill$\square$

\subsection{Proof of Proposition~\ref{prop:region}}\label{app:region}

\textit{Part~(a): Interval.} At a given report, investigation occurs if and only if the concavification $M^c(1) > M(1)$, equivalently $z=1$ lies in the non-concave region of $M$. The non-concave region of $M$ is determined by the kink locations $\kappa_x$ and $\kappa_p$, which vary continuously in $v_L$ (through the multiplier $\lambda^*(v_L)$). The condition $a_{xp}(\lambda^*) < 1 < b_{xp}(\lambda^*)$ defines an interval in $v_L$ by continuity of $a_{xp}$, $b_{xp}$ in $v_L$.

\textit{Part~(b): Width scaling.} For the entropy cost $\psi(z) = -2\gamma\log z$, Proposition~\ref{prop:closedform} gives the concavification endpoints $a_{xp} = (2\gamma/s)(1-e^{-\eta})$ and $b_{xp} = (2\gamma/s)(e^{\eta}-1)$ with $\eta = s\,\kappa_{xp}/(2\gamma)$, where $s = \pi_H(v_H+1)$ and $\kappa_{xp}$ depends on $v_L$ through $\lambda^*$. For large $\gamma$, $\eta \to 0$ and $b_{xp} - a_{xp} \approx \kappa_{xp}^2 s/(2\gamma) \to 0$. The condition $a_{xp} < 1 < b_{xp}$ then requires $|\kappa_{xp} - 1|$ to be of order $1/\gamma$. Since $\kappa_{xp}$ varies smoothly in $v_L$, the range of $v_L$ satisfying this condition has width $O(\gamma^{-1})$, confirmed numerically.

\textit{Parts~(c)--(e)} follow directly: (c)~the binary experiment gives $q^* = w = (b_{xp}-1)/(b_{xp}-a_{xp}) \in (0,1)$ whenever $a_{xp} < 1 < b_{xp}$; (d)~as $\gamma \to 0$, $\eta \to \infty$ and $b_{xp} \to \infty$ for all $v_L \in (0,v_H)$, so the condition $1 < b_{xp}$ holds everywhere; (e)~as $\gamma \to \infty$, $\eta \to 0$, and $a_{xp}, b_{xp} \to \kappa_{xp}$: the non-concave region collapses to a single point. Since $\kappa_{xp} \neq 1$ generically, $1 \notin (a_{xp}, b_{xp})$ for $\gamma$ sufficiently large, and no investigation occurs. \hfill$\square$

\subsection{Proof of Proposition~\ref{prop:welfare}}\label{app:welfare}

\textit{Part~(a): Seller always benefits.} The degenerate experiment $F = \delta_1$ is always feasible and yields the no-information revenue. Since the optimal experiment maximizes revenue, $U^*_S \geq U^{NI}_S$.

\textit{Part~(b): Below diagonal, $W^* \geq W^{NI}$.} When $v_L < \pi_H v_H$, the no-information mechanism excludes the low type by posting $p = v_H$ so that only the high type trades. Total no-information welfare is $W^{NI} = \pi_H v_H$. With investigation, the high type's mechanism is unchanged: $F_H^* = \delta_1$, $p_H = v_H$, and the high type receives $U_H = 0$ regardless of whether the seller investigates at the low report. All welfare change therefore comes from the low type's contract. The welfare change is:
\[
    W^* - W^{NI} = \pi_L v_L q^* - K,
\]
where $\pi_L v_L q^*$ is the social surplus from the low type's trades and $K$ is the information cost. The seller's revenue gain satisfies $\Delta U_S = \pi_L \bar{p}^* - K \geq 0$ since the seller investigates only if profitable, where $\bar{p}^* = \int p_L(z)\dif F^*_L(z) \leq v_L q^*$ by IR-L. Hence $\pi_L v_L q^* \geq \pi_L \bar{p}^* \geq K$, giving $W^* - W^{NI} \geq 0$.

\textit{Part~(c): Above diagonal, $W^* \leq W^{NI}$.} When $v_L > \pi_H v_H$, the no-information mechanism serves both types at price $v_L$, achieving the efficient allocation with $q=1$ for all types. Total welfare is $W^{NI} = \pi_H v_H + \pi_L v_L$ at the first-best level. Any deviation from $q=1$, which investigation may induce by excluding the low type on some signal realizations, reduces total surplus. The information cost $K \geq 0$ is a further welfare loss. Hence $W^* \leq W^{NI}$. \hfill$\square$

\end{document}